\documentclass[fontsize=10pt]{scrartcl}
\usepackage[utf8]{inputenc}
\usepackage[T1]{fontenc}
\usepackage[american]{babel}
\usepackage[autostyle,english=american]{csquotes}
\usepackage[theoremfont]{newpxtext}
\usepackage{newpxmath}

\usepackage{amsthm}
\usepackage{amssymb}
\usepackage{amsmath}
\usepackage{enumitem}
\usepackage{mathtools}
\usepackage[headsepline]{scrlayer-scrpage}
\usepackage[pdfborder={0 0 0.5},pdfpagelabels,colorlinks,linkcolor=blue]{hyperref}
\usepackage{cleveref}

\newtheorem{thm}{Theorem}
\newtheorem{lma}[thm]{Lemma}

\theoremstyle{definition}
\newtheorem{defn}[thm]{Definition}
\newtheorem{rem}[thm]{Remark}
\newtheorem{cond}[thm]{Condition}
\newtheorem{defrem}[thm]{Definition and Remark}
\newlist{thmlist}{enumerate}{1}
\setlist[thmlist]{label=(\roman{thmlisti}), ref=\thethm.(\roman{thmlisti}),noitemsep}
\newlist{thmoptionlist}{enumerate}{1}
\setlist[thmoptionlist]{label=(\alph{thmoptionlisti}), ref=\thethm.(\roman{thmlisti}).(\alph{thmoptionlisti}),noitemsep}
\newlist{lmalist}{enumerate}{1}
\setlist[lmalist]{label=(\roman{lmalisti}), ref=\thethm.(\roman{lmalisti}),noitemsep}
\newlist{proplist}{enumerate}{1}
\setlist[proplist]{label=(\roman{proplisti}), ref=\thethm.(\roman{proplisti}),noitemsep}
\newlist{corlist}{enumerate}{1}
\setlist[corlist]{label=(\roman{corlisti}), ref=\thethm.(\roman{corlisti}),noitemsep}
\newlist{deflist}{enumerate}{1}
\setlist[deflist]{label=(\roman{deflisti}), ref=\thethm.(\roman{deflisti}),noitemsep}
\newlist{condlist}{enumerate}{1}
\setlist[condlist]{label=(\roman{condlisti}), ref=\thethm.(\roman{condlisti}),noitemsep}
\newlist{condoptionlist}{enumerate}{1}
\setlist[condoptionlist]{label=(\arabic{condoptionlisti}), ref=\thethm.(\roman{condlisti}).(\arabic{condoptionlisti}),noitemsep}
\numberwithin{thm}{section}
\allowdisplaybreaks
\crefname{chapter}{Chapter}{Chapters}
\crefname{section}{Section}{Sections}
\crefname{subsection}{Section}{Sections}
\crefname{thm}{Theorem}{Theorems}
\crefname{lma}{Lemma}{Lemmas}
\crefname{prop}{Proposition}{Propositions}
\crefname{cor}{Corollary}{Corollaries}
\crefname{defn}{Definition}{Definitions}
\crefname{rem}{Remark}{Remarks}
\crefname{cond}{Condition}{Conditions}

\crefname{equation}{}{}
\crefname{thmlisti}{Theorem}{Theorem}
\crefname{thmoptionlisti}{option}{options}
\crefname{lmalisti}{Lemma}{Lemma}
\crefname{proplisti}{Proposition}{Proposition}
\crefname{corlisti}{Corollary}{Corollary}
\crefname{deflisti}{Definition}{Definition}
\crefname{condlisti}{Condition}{Condition}
\crefname{condoptionlisti}{case}{cases}
\crefname{equation}{}{}
\newcommand{\f}{f^\alpha}
\newcommand{\g}{g^\alpha}

\newcommand{\e}{q_\alpha}

\newcommand{\m}{m_\alpha}
\renewcommand{\v}{\widehat{v}_\alpha}
\newcommand{\vo}{v_\alpha^0}

\newcommand{\etaa}{\eta^\alpha}
\newcommand{\E}{\mathcal E^\alpha}
\newcommand{\F}{\mathcal F^\alpha}
\newcommand{\G}{\mathcal G^\alpha}
\newcommand{\R}{\mathbb{R}}
\newcommand{\N}{\mathbb{N}}

\newcommand{\M}{\mathcal M}
\newcommand{\suma}{\sum_{\alpha=1}^N}

\let\originalleft\left
\let\originalright\right
\renewcommand{\left}{\mathopen{}\mathclose\bgroup\originalleft}
\renewcommand{\right}{\aftergroup\egroup\originalright}

\let\div\relax
\DeclareMathOperator{\div}{div}
\DeclareMathOperator{\curl}{curl}

\newcommand{\ext}{\mathrm{ext}}
\newcommand{\tot}{\mathrm{tot}}

\makeatletter
\def\env@cases{\let\@ifnextchar\new@ifnextchar\left\lbrace\def\arraystretch{1.2}\array{@{}r@{\quad}l@{}}}
\makeatother
\automark{section}
\automark*{subsection}
\ihead{\leftmark}
\ohead{\thepage}
\begin{document}

\title{Confined Steady States of a Relativistic Vlasov-Maxwell Plasma in a Long Cylinder}
\author{Jörg Weber\\ \textit{University of Bayreuth, 95440 Bayreuth, Bavaria, Germany}\\ \texttt{Joerg.Weber@uni-bayreuth.de}}
\date{}
\maketitle
\begin{abstract}
	The time evolution of a collisionless plasma is modeled by the relativistic Vlasov-Maxwell system which couples the Vlasov equation (the transport equation) with the Maxwell equations of electrodynamics. In this work, the setting is two and one-half dimensional, that is, the distribution functions of the particles species are independent of the third space dimension. We consider the case that the plasma is located in an infinitely long cylinder and is influenced by an external magnetic field. We prove existence of stationary solutions and give conditions on the external magnetic field under which the plasma is confined inside the cylinder, i.e., it stays away from the boundary of the cylinder.
	
	\vspace*{3mm}
	
	
	\vspace*{3mm}
	
\end{abstract}

\numberwithin{equation}{section}
\section{Introduction}
If a plasma is sufficiently rarefied or hot, collisions among the plasma particles can be neglected and the time evolution of this plasma can be modeled by the relativistic Vlasov-Maxwell system. We consider the case that the plasma is contained in some open set $\Omega\subset\R^3$ and that the particles and electromagnetic fields, respectively, are subject to purely reflecting and perfect conductor boundary conditions, respectively. In particular, the system reads
\begin{subequations}\label{eq:WholeSystem}
	\begin{align}
	\partial_t\f+\v\cdot\partial_x\f+\e\left(E+\v\times B^\tot\right)\cdot\partial_v\f&=0&\ \mathrm{on}\ \left[0,T\right]\times\Omega\times\R^3,\label{eq:WholeVl}\\
	\f_-&=K\f_+&\ \mathrm{on}\ \gamma_T^-,\label{eq:WholeBoun}\\
	\f\left(0\right)&=\mathring\f&\ \mathrm{on}\ \Omega\times\R^3,\label{eq:WholeInitf}\\
	\partial_tE-\curl_xB&=-4\pi j&\ \mathrm{on}\ \left[0,T\right]\times\Omega,\label{eq:WholeMax1}\\
	\partial_tB+\curl_xE&=0&\ \mathrm{on}\ \left[0,T\right]\times\Omega,\label{eq:WholeMax2}\\
	\div_xE&=4\pi\rho&\ \mathrm{on}\ \left[0,T\right]\times\Omega,\label{eq:WholeMax3}\\
	\div_xB&=0&\ \mathrm{on}\ \left[0,T\right]\times\Omega,\label{eq:WholeMax4}\\
	E\times n=B^\tot\cdot n&=0&\ \mathrm{on}\ \left[0,T\right]\times\partial\Omega,\label{eq:WholeMaxBoun}\\
	\left(E,B\right)\left(0\right)&=\left(\mathring E,\mathring B\right)&\ \mathrm{on}\ \Omega.\label{eq:WholeInitEH}
	\end{align}
\end{subequations}
This set of equations, imposed on some time interval $\left[0,T\right]$, describes the time evolution of a collisionless plasma which consists of $N$ particle species. Equations \cref{eq:WholeVl,eq:WholeBoun,eq:WholeInitf} are to hold for each $\alpha=1,\dots,N$, where \cref{eq:WholeVl} is the Vlasov equation for the density $\f=\f\left(t,x,v\right)$ of the $\alpha$-th particle species. These densities depend on time $t\in\left[0,T\right]$, position $x\in\Omega$ and momentum $v\in\R^3$, from which the relativistic velocity is computed via
\begin{align*}
\v=\frac{v}{\sqrt{m_\alpha^2+\left|v\right|^2}}.
\end{align*}
Here and throughout this paper, $\left|\cdot\right|$ denotes the Euclidean norm. The quantities $m_\alpha$ and $\e$ are the rest mass and charge of a particle of the $\alpha$-th species.

Equation \cref{eq:WholeInitf} is the initial condition for $\f$ and \cref{eq:WholeBoun} describes the boundary condition on $\partial\Omega$. Here, $\f_\pm$ are the restrictions of $\f$ to
\begin{align*}
\gamma_{T}^\pm&\coloneqq\left\{\left(t,x,v\right)\in\left[0,T\right]\times\partial\Omega\times\R^3\mid v\cdot n\left(x\right)\gtrless 0\right\},
\end{align*}
The operator $K$ describes pure reflection on $\partial\Omega$ via
\begin{align*}
\left(Kh\right)\left(t,x,v\right)=h\left(t,x,v-2\left(v\cdot n\left(x\right)\right)\right),
\end{align*}
Above, $n\left(x\right)$ denotes the outer unit normal of $\partial\Omega$ at $x\in\partial\Omega$.

Equations \cref{eq:WholeMax1,eq:WholeMax2,eq:WholeMax3,eq:WholeMax4} are the Maxwell equations for the electromagnetic fields $E=E\left(t,x\right)$, $B=B\left(t,x\right)$ with initial condition \cref{eq:WholeInitEH}. The source terms are
\begin{align*}
j\coloneqq\suma\e\int_{\R^3}\v\f\,dv,\quad\rho\coloneqq\suma\e\int_{\R^3}\f\,dv,
\end{align*}
the current and charge density $j$ and $\rho$ induced by the plasma particles. Moreover, \cref{eq:WholeMaxBoun} is the perfect conductor boundary condition.

Furthermore, we consider the case that an external magnetic field $B^\ext$ influences the plasma particles. Accordingly, the total magnetic field $B^\tot=B+B^\ext$ appears in the Lorentz force in \cref{eq:WholeVl}.

The aim of this paper is to answer the following two questions: First, for given time-independent external magnetic field, is there a stationary solution of \cref{eq:WholeSystem}? Second, are there stationary solutions that are confined in $\Omega$, i.e., the particles stay away from the boundary of their container, if the external magnetic field is adjusted suitably? 

Before we analyze these problems, we first discuss the basic ideas for plasma confinement -- more information on fusion plasma physics can be found in the classical book of Stacey \cite{Sta12}. The physical basis for confinement is the fact that charged particles spiral about magnetic field lines. The so called gyroradius, that is, the radius of such a spiral, is inversely proportional to the strength of the magnetic field. This gives rise to the idea of linear confinement devices: The fusion reactor is a long cylinder and the external magnetic field points in the direction of the symmetry axis of this cylinder. If this external magnetic field is sufficiently strong, the gyroradii of the plasma particles will be smaller than radius of the cylinder, whence the plasma is confined in the fusion device. However, this setting cannot prevent the plasma current from having a non-vanishing component in the direction of the symmetry axis. Thus, there will be losses at the ends of the long cylinder. In practice, one can try to overcome this problem by one of the two following modifications: First, so-called magnetic mirrors are added at these ends. Second, the long cylinder is bent into a torus. This second idea is pursued typically in modern research. Toroidal geometry has the advantage of avoiding such losses, but has the disadvantage that it gives rise to drifts of the plasma particles, which finally cause the particles moving radially outwards and thus make confinement impossible. Therefore, the external magnetic field needs to have a poloidal component additional to its toroidal one. This approach then leads to Tokamak devices.

However, analyzing the problem of existence of confined steady states from a mathematics point of view in toroidal geometry seems quite hard. As a first step towards this, we consider the set-up of a linear confinement device instead. For mathematical reasons, it will be convenient to assume that the cylinder is infinitely long (which is of course not conceivable from a practical point of view). Thus, we fix $R_0>0$ and let
\begin{align*}
\Omega\coloneqq\left\{x\in\R^3\mid x_1^2+x_2^2<R_0^2\right\}.
\end{align*}
Because of the axial symmetry of the set-up, it is natural to work with cylindrical coordinates $\left(r,\varphi,x_3\right)$. In these coordinates, we simply have $\Omega=\left\{x\in\R^3\mid r<R_0\right\}$.

In the following, there often occur cylindrical coordinates and the corresponding local, orthonormal coordinate basis $\left(e_r,e_\varphi,e_3\right)$, where
\begin{align*}
e_r=\left(\cos\varphi,\sin\varphi,0\right),\quad e_\varphi=\left(-\sin\varphi,\cos\varphi,0\right),\quad e_3=\left(0,0,1\right).
\end{align*}
For a vector $w\in\R^3$ we denote with $w_r$, $w_\varphi$, and $w_3$ the coordinates of $w$ in this local coordinate system, i.e.,
\begin{align*}
w_r=w\cdot e_r,\quad w_\varphi=w\cdot e_\varphi,\quad w_3=w\cdot e_3.
\end{align*}
Note that the perfect conductor boundary condition $E\times n=0=B\cdot n$ now reduces to $E_\varphi=E_3=B_r^\tot=0$ in the case of $\Omega$ being an infinitely long cylinder, since here $n=e_r$.

It is convenient to introduce electromagnetic potentials, which will be the functions we work with mostly, namely the electric scalar potential $\phi$ and the magnetic vector potential $A^\tot=A+A^\ext$, which splits into the internal and external potentials $A$ and $A^\ext$. The electromagnetic fields and potentials are related via
\begin{align}\label{eq:potentialsfields}
E=-\partial_x\phi-\partial_tA^\tot,\quad B=\curl_xA,\quad B^\ext=\curl_xA^\ext.
\end{align}
Then, Gauss's law for magnetism ($\div_xB=0$) and Faraday's law ($\partial_tB+\curl_xE=0$) are automatically satisfied. There is some freedom to demand a certain gauge condition on the potentials. We will consider Lorenz gauge for the internal potentials
\begin{align}\label{eq:Lorenzgauge}
\partial_t\phi+\div_xA=0,
\end{align}
which of course is the same as Coulomb gauge
\begin{align*}
\div_xA=0
\end{align*}
if the potentials are independent of time, and similarly $\div_xA^\ext=0$ for the external potential.

Similar set-ups have already been studied earlier, for example in \cite{Pou92,Rei92}. The basic strategy to obtain steady states was first mentioned in \cite{Deg90}. Closely related to our considerations is \cite{BF93}, where (among other set-ups) existence of steady states in an infinitely long cylinder without external magnetic field was proved. However, an important condition there is that there is only one particle species and thus only a fixed sign of particle charges appears. Therefore, $\rho$ has a fixed sign and $\phi$ is monotone, which is crucial for the considerations in \cite{BF93}. As opposed to this, we allow positively and negatively charged particles.

The question about existence of confined steady states for a Vlasov-Poisson plasma (that is, $B=0$) by means of an external magnetic field was considered in \cite{Sku14} and \cite{Kno19}. The approach of the latter work is similar to ours but needs some smallness assumption on the ansatz functions, which we can avoid, and is restricted to homogeneous external magnetic fields parallel to the symmetry axis. Also, we refer to \cite{CCM12,CCM14,CCM15,CCM16} for considerations about confinement of a Vlasov-Poisson plasma.

There are also some papers concerning Vlasov-Maxwell plasmas and the problem of their confinement as well as concerning their stability \cite{Han10,NNS15,NS14,Zha19,ZHR15}.

Another approach to control a plasma by means of external fields has been pursued by Knopf and the author in \cite{Kno18,KW18,Web18,Web191,Web192}.

This work is organized as follows: In \cref{sec:SymmConser}, we state some basic assumptions on the symmetry of the appearing functions and state the corresponding invariant quantities, which lead to the natural ansatz concerning the densities $\f$. This ansatz, together with a basic definition and some useful preliminary lemmas and tools, is the content of \cref{sec:ststdefansatz}. In \cref{sec:existence,sec:confined}, we answer the above-mentioned questions. In particular, we prove existence of a steady state for a given external magnetic field and give conditions on the external magnetic potential under which the steady state is confined.

\section{Symmetries and invariants}\label{sec:SymmConser}
Due to the symmetry properties of $\Omega$, it is natural to consider the case that the tuple $\left(\left(\f\right)_\alpha,\phi,A,A^\ext\right)$ has some symmetry properties as well:

Firstly, as $\Omega$ is invariant under translations in the $e_3$-direction, we assume that the tuple $\left(\left(\f\right)_\alpha,\phi,A,A^\ext\right)$ is independent of $x_3$, that is,
\begin{align*}
\f=\f\left(t,x_1,x_2,v_1,v_2,v_3\right),\; \phi=\phi\left(t,x_1,x_2\right),\; A=A\left(t,x_1,x_2\right),\; A^\ext=A^\ext\left(t,x_1,x_2\right). 
\end{align*}
Then, of course the same property also holds for $E$, $B$, and $B^\ext$. With this assumption, the resulting system is also called the \enquote{two and one-half dimensional} relativistic Vlasov-Maxwell system, since an $\f$ as above only depends on two space and three momentum variables. Due to Glassey and Schaeffer \cite{GS97}, unique, classical solutions to the resulting system without boundary conditions on $\partial\Omega$ and with $B^\ext=0$ exist globally in time under suitable assumptions about the initial data.

Secondly, as $\Omega$ is invariant under rotations about the $x_3$-axis, we assume that the tuple $\left(\left(\f\right)_\alpha,\phi,A,A^\ext\right)$ has the following property:
\begin{align*}
\f\left(t,Rx,Rv\right)=\f\left(t,x,v\right),\;\phi\left(t,Rx\right)=\phi\left(t,x\right),\;A\left(t,Rx\right)=RA\left(t,x\right),\;A^\ext\left(t,Rx\right)=RA^\ext\left(t,x\right)
\end{align*}
for any rotation $R\in\R^{3\times 3}$ about the $x_3$-axis. With the use of cylindrical coordinates, this assumption about the potentials is equivalent to the assumption that
\begin{align*}
\phi=\phi\left(t,r,x_3\right)
\end{align*}
and that the components of the vector potentials in the local coordinate basis $\left(e_r,e_\varphi,e_3\right)$ be independent of the angle $\varphi$, that is,
\begin{align*}
A_r&=A_r\left(t,r,x_3\right),&A_\varphi&=A_\varphi\left(t,r,x_3\right),&A_3&=A_3\left(t,r,x_3\right),\\
A_r^\ext&=A_r^\ext\left(t,r,x_3\right),&A_\varphi^\ext&=A_\varphi^\ext\left(t,r,x_3\right),&A_3^\ext&=A_3^\ext\left(t,r,x_3\right).
\end{align*}
With this symmetry, we can also reduce the number of variables in $\left(x,v\right)$-space from six to five and can write $f=f\left(r,x_3,\theta,u,v_3\right)$ where $u=\sqrt{v_1^2+v_2^2}$ and $\theta$ is the angle between $\left(x_1,x_2\right)$ and $\left(v_1,v_2\right)$. However, we will not make use of the Vlasov equation written in these variables.

Additionally to these two space symmetries, we consider time symmetry, i.e., the tuple $\left(\left(\f\right)_\alpha,\phi,A,A^\ext\right)$ is assumed to be independent of $t$, since we are interested in the existence of (confined) steady states.

In cylindrical coordinates, there holds (for any scalar function $\phi$ and any vector-valued function $A$)
\begin{align*}
\partial_x\phi&=e_r\partial_r\phi+\frac{1}{r}e_\varphi\partial_\varphi\phi+e_3\partial_{x_3}\phi,\\
\curl_xA&=e_r\left(\frac{1}{r}\partial_\varphi A_3-\partial_{x_3}A_\varphi\right)+e_\varphi\left(\partial_{x_3}A_r-\partial_rA_3\right)+\frac{1}{r}e_3\left(\partial_r\left(rA_\varphi\right)-\partial_\varphi A_r\right).
\end{align*}
Thus, assuming time symmetry and the two space symmetries, \cref{eq:potentialsfields} becomes
\begin{gather*}
E_r=-\partial_r\phi,\quad E_\varphi=E_3=0,\\
B_r=0,\quad B_\varphi=-\partial_rA_3,\quad B_3=\frac{1}{r}\partial_r\left(rA_\varphi\right),\\
B_r^\ext=0,\quad B_\varphi^\ext=-\partial_rA_3^\ext,\quad B_3^\ext=\frac{1}{r}\partial_r\left(rA_\varphi^\ext\right).
\end{gather*}
Hence, perfect conductor boundary conditions on $\partial\Omega$ are always satisfied in this case and we can let $A_r=0$ without loss of generality since $A_r$ does not affect the electromagnetic fields.

Using the gauge \cref{eq:Lorenzgauge}, the remaining Maxwell's equations, i.e., $\partial_tE-\curl_xB=-4\pi j$ and $\div_xE=4\pi\rho$, become
\begin{align}\label{eq:Maxwellpotentials}
\partial_t^2\phi-\Delta_x\phi=4\pi\rho,\qquad\partial_t^2A-\Delta_xA=4\pi j,
\end{align}
where the latter equation is to be understood componentwise (in Cartesian coordinates). In cylindrical coordinates, we have (for any scalar function $\phi$ and any vector-valued function $A$)
\begin{align*}
\Delta_x\phi&=\frac{1}{r}\partial_r\left(r\partial_r\phi\right)+\frac{1}{r^2}\partial_\varphi^2\phi+\partial_{x_3}^2\phi,\\
\Delta_xA&=e_r\left(\Delta_xA_r-\frac{1}{r^2}A_r-\frac{2}{r^2}\partial_\varphi A_\varphi\right)+e_\varphi\left(\Delta_xA_\varphi-\frac{1}{r^2}A_\varphi+\frac{2}{r^2}\partial_\varphi A_r\right)+e_3\Delta_xA_3.
\end{align*}
Thus, assuming time symmetry, the two space symmetries, and $A_r=0$, on the one hand the gauge \cref{eq:Lorenzgauge} is automatically satisfied, as there holds
\begin{align}\label{eq:divcylcoor}
\div_xA=\frac{1}{r}\partial_r\left(rA_r\right)+\frac{1}{r}\partial_\varphi A_\varphi+\partial_{x_3}A_3
\end{align}
in general, and on the other hand \cref{eq:Maxwellpotentials} becomes
\begin{align}\label{eq:ODEpotentials}
-\frac{1}{r}\left(r\phi'\right)'=4\pi\rho,\qquad-\left(\frac{1}{r}\left(rA_\varphi\right)'\right)'=4\pi j_\varphi,\qquad-\frac{1}{r}\left(rA_3'\right)'=4\pi j_3.
\end{align}
As $\phi$, $A_\varphi$, and $A_3$ only depend on $r$, we denote the $r$-derivative with simply $'$. Note that the choice $A_r=0$ launches the constraint
\begin{align*}
j_r=0,
\end{align*}
i.e., no radial currents are allowed to appear.

A basic physical principle is that to each symmetry there corresponds an invariant. For each of the two space symmetries, we can derive an invariant from the Lagrangian (without the use of any gauge)
\begin{align*}
\mathcal L^\alpha=\mathcal L^\alpha\left(t,x,\dot x\right)=-\sqrt{1-\left|\dot x\right|^2}-\e\left(\phi\left(t,x\right)-\dot x\cdot A^\tot\left(t,x\right)\right).
\end{align*}
In particular, the invariant
\begin{align*}
\G\coloneqq\partial_{\dot{x_3}}\mathcal L^\alpha=v_3+\e A_3^\tot
\end{align*}
corresponds to translation invariance and
\begin{align*}
\F\coloneqq\partial_{\dot\varphi}\mathcal L^\alpha=r\left(v_\varphi+\e A_\varphi^\tot\right)
\end{align*}
corresponds to rotational symmetry. Note that $\F$ (the \enquote{canonical angular momentum}) and $\G$ are the $\varphi$- and the third component of the so-called \enquote{canonical momentum}
\begin{align*}
p_\alpha=v+\e A^\tot.
\end{align*}
In the variables $\left(x,p_\alpha\right)$, the particle energy
\begin{align*}
\E\coloneqq\vo+\e\phi\coloneqq\sqrt{\m^2+\left|v\right|^2}+\e\phi=\sqrt{\m^2+\left|p_\alpha-\e A^\tot\right|^2}+\e\phi
\end{align*}
is the (in general time-dependent) Hamiltonian governing the motion of the particles of the $\alpha$-th species. Assuming that the electromagnetic potentials are independent of time, $\E$ is also independent of time and thus another invariant, the one corresponding to time symmetry.

\section{Steady states -- Definition and ansatz}\label{sec:ststdefansatz}
The preceding considerations about symmetry motivate the definition of what we call a (confined) steady state in our set-up. Before that we collect our symmetry assumptions:
\begin{defrem}\phantom{}
	\begin{enumerate}[label=(\alph*),leftmargin=*]
		\item A function $f\colon\overline\Omega\to\R$ / a function $\phi\colon\overline\Omega\to\R$ / a vector field $A\colon\overline\Omega\to\R^3$ is called
		\begin{enumerate}[label=(\roman*)]
			\item independent of $x_3$ if $\partial_{x_3}f=0$ / $\partial_{x_3}\phi=0$ / $\partial_{x_3}A=0$;
			\item axially symmetric if $f\left(Rx,Rv\right)=f\left(x,v\right)$ for any $x\in\overline\Omega$, $v\in\R^3$, and rotation $R\in\R^{3\times 3}$ about the $x_3$-axis / $\phi\left(Rx\right)=\phi\left(x\right)$ for any $x\in\overline\Omega$ and rotation $R\in\R^{3\times 3}$ about the $x_3$-axis / $A\left(Rx\right)=RA\left(x\right)$ for any $x\in\overline\Omega$ and rotation $R\in\R^{3\times 3}$ about the $x_3$-axis.
		\end{enumerate}
		\item With these two symmetries, the functions $\phi$, $A_r$, $A_\varphi$, and $A_3$ only depend on $r$. Accordingly, we will often view them as functions on $\left[0,R_0\right]$.
		\item An axially symmetric vector field $A$ automatically satisfies $A_1\left(x\right)=A_2\left(x\right)=0$ if $x_1=x_2=0$, i.e., if $x$ lies on the $x_3$-axis.
	\end{enumerate}
\end{defrem}
We proceed with an assumption about the external potential, which is supposed to hold henceforth:
\begin{cond}\label{cond:externalpotential}
	The external potential $A^\ext\colon\overline\Omega\to\R$ is independent of $x_3$ and axially symmetric such that $A_r^\ext=0$ and $A_\varphi^\ext,A_3^\ext\in C^1\left(\left[0,R_0\right]\right)$ (viewed as functions of $r$) with $A_\varphi^\ext\left(0\right)=A_3^\ext\left(0\right)=\left(A_3^\ext\right)'\left(0\right)=0$.
\end{cond}
Note that $A_3^\ext\left(0\right)=0$ can be assumed -- for simplicity -- without loss of generality, since adding a constant to $A_3^\ext$ does not affect $B^\ext$ because of $\curl_xe_3=0$ (as opposed to this, this invariance under adding constants does not hold for $A_\varphi^\ext$, as $\curl_xe_\varphi\neq 0$).

We first prove some technicalities:
\begin{lma}\label{lma:regularitypotentials}
	Let $\phi, A_\varphi, A_3\in C^1\left(\left[0,R_0\right]\right)$ with
	\begin{align}\label{eq:potentialsat0}
	\phi'\left(0\right)=A_\varphi\left(0\right)=A_3'\left(0\right)=0
	\end{align}
	and assume $A_r=0$. Then there holds:
	\begin{lmalist}
		\item\label{lma:regularitypotentialsi} The potentials $\phi=\phi\left(x\right)$ and $A=A\left(x\right)$ are continuously differentiable on $\overline\Omega$. Thus, the electromagnetic fields
		\begin{align}\label{eq:EBtot}
		E=-\partial_x\phi=-\phi'e_r,\quad B=\curl_xA=-A_3'e_\varphi+\frac{1}{r}\left(rA_\varphi\right)'e_3
		\end{align}
		are continuous on $\overline\Omega$. Moreover, $\div_xA=0$ on $\overline\Omega$.
		\item\label{lma:regularitypotentialsii} If $\phi, A_3\in C^2\left(\left[0,R_0\right]\right)$, they are even twice continuously differentiable on $\overline\Omega$ with respect to $x$. Accordingly, $E$ is of class $C^1$ on $\overline\Omega$. If moreover $A_\varphi\in C^2\left(\left]0,R_0\right]\right)$ such that
		\begin{align}\label{eq:Aphilimitr0}
		A_\varphi'\left(r\right)-\frac{A_\varphi\left(r\right)}{r}=\mathcal O\left(r\right),\quad A_\varphi''\left(r\right)=\mathcal O\left(1\right)\quad\mbox{for }r\to 0,
		\end{align}
		then $A\in W^{2,\infty}\left(\Omega;\R^3\right)\cap C^2\left(\overline\Omega\setminus\R e_3;\R^3\right)$. Accordingly, $B$ is of class $W^{1,\infty}$ on $\Omega$ and of class $C^1$ on $\overline\Omega\setminus\R e_3$.
	\end{lmalist}
	\begin{proof}
		We easily see that the maps $x\mapsto\phi\left(x\right)$ and $x\mapsto A_3\left(x\right)e_3$ are (twice) continuously differentiable on $\overline\Omega$ if the maps $r\mapsto\phi\left(r\right)$ and $r\mapsto A_3\left(r\right)$ are (twice) continuously differentiable on $\left[0,R_0\right]$, since $\phi'\left(0\right)=\left(A_3^\tot\right)'\left(0\right)=0$. There remains to take care of $x\mapsto A_\varphi\left(x\right)e_\varphi\left(x\right)$, in particular at $r=0$. Indeed, this map can be continuously extended to whole $\overline\Omega$ because of $A_\varphi\left(0\right)=0$ and is differentiable for $r>0$ with
		\begin{align}
		\partial_x\left(A_\varphi e_\varphi\right)\left(r,\varphi\right)=\begin{pmatrix}\sin\varphi\cos\varphi\left(-A_\varphi'\left(r\right)+\frac{A_\varphi\left(r\right)}{r}\right)&-\sin^2\varphi\left(A_\varphi'\left(r\right)-\frac{A_\varphi\left(r\right)}{r}\right)-\frac{A_\varphi\left(r\right)}{r}&0\\\cos^2\varphi\left(A_\varphi'\left(r\right)-\frac{A_\varphi\left(r\right)}{r}\right)+\frac{A_\varphi\left(r\right)}{r}&\sin\varphi\cos\varphi\left(A_\varphi'\left(r\right)-\frac{A_\varphi\left(r\right)}{r}\right)&0\\0&0&0\end{pmatrix}\label{eq:partialxAephi}
		\end{align}
		and all entries have a limit as $r\to 0$. Hence, also $A_\varphi e_\varphi$ is continuously differentiable on whole $\overline\Omega$. Furthermore, $A$ is divergence free with respect to $x$, as was already observed in \cref{sec:SymmConser} because of \cref{eq:divcylcoor}. Thus, \ref{lma:regularitypotentialsi} is proved. If moreover the assumptions about $A_\varphi$ in \ref{lma:regularitypotentialsii} are satisfied, second order derivatives (with respect to $x$) of $A_\varphi e_\varphi$ are bounded for $r\to 0$, since we see by differentiating the entries of \cref{eq:partialxAephi} once more that these second order derivatives are expressions in $\sin\varphi$, $\cos\varphi$, $\frac{1}{r}\left(A_\varphi'\left(r\right)-\frac{A_\varphi\left(r\right)}{r}\right)$, and $A_\varphi''\left(r\right)$, and thus bounded by assumption. Therefore, all second order derivatives exist on $\Omega$ in the weak sense, coincide with the classical derivatives almost everywhere, and are bounded. This proves the remaining part of \ref{lma:regularitypotentialsii}.
	\end{proof}
\end{lma}
Note that under \cref{cond:externalpotential} the external potential $A^\ext$ is continuously differentiable on $\overline\Omega$ and divergence free. Also, the external magnetic field $B^\ext=\curl_xA^\ext$ is continuous on $\overline\Omega$.
\begin{rem}
	In \cref{lma:regularitypotentialsii}, we cannot expect that $A\in C^2\left(\overline\Omega;\R^3\right)$ in general if $A_\varphi\in C^2\left(\left[0,R_0\right]\right)$ and \cref{eq:Aphilimitr0} holds, as the example $A_\varphi\left(r\right)=r^2$ shows, since
	\begin{align*}
	\Delta_x\left(A_\varphi e_\varphi\right)_1=-\Delta_x\left(r^2\sin\varphi\right)=-3\sin\varphi
	\end{align*}
	has no limit for $r\to 0$.
\end{rem}
We proceed with a basic definition:
\begin{defn}\label{def:steadystate}
	Let \cref{cond:externalpotential} hold.
	\begin{enumerate}[label=(\alph*),leftmargin=*]
		\item A tuple $\left(\left(\f\right)_\alpha,\phi,A\right)$ is called an axially symmetric steady state of the two and one-half dimensional relativistic Vlasov-Maxwell system on $\overline\Omega$ with external potential $A^\ext$ (hereafter abbreviated as steady state) if the following conditions are satisfied:
		\begin{enumerate}[label=(\roman*)]
			\item For each $\alpha=1,\dots,N$, the functions $\f\colon\overline\Omega\to\R^3\to\left[0,\infty\right[$ are continuously differentiable satisfying $\f\left(x,\cdot\right)\in L^1\left(\R^3\right)$ for each $x\in\overline\Omega$.
			\item The potentials satisfy
			\begin{align*}
			\phi\in C^2\left(\overline\Omega\right),\quad A\in C^1\left(\overline\Omega;\R^3\right)\cap C^2\left(\overline\Omega\setminus\R e_3;\R^3\right)\cap W^{2,\infty}\left(\Omega;\R^3\right).
			\end{align*}
			(This condition is motivated in view of \cref{lma:regularitypotentials}.)
			\item Any $\f$ and $\phi$, $A$ are independent of $x_3$ and axially symmetric.
			\item The equations
			\begin{subequations}
				\begin{align}
				\v\cdot\partial_x\f+\e\left(E+\v\times B^\tot\right)\cdot\partial_v\f=0&&\ \mathrm{on}\ \overline\Omega\times\R^3,\label{eq:Vlstat}\\
				\f\left(x,v-2v_re_r\right)=f\left(x,v\right),&& x\in\partial\Omega,v\in\R^3,v_r<0,\label{eq:BCstat}\\
				-\Delta_x\phi=4\pi\rho,\quad-\Delta_xA=4\pi j,\quad\div_xA=0&&\ \mathrm{on}\ \overline\Omega,
				\end{align}
			\end{subequations}
			are satisfied. Here, $e_r=e_r\left(x\right)$, $v_r=v\cdot e_r$, and
			\begin{gather*}
			E=-\partial_x\phi,\;B^\tot=\curl_x\left(A+A^\ext\right),\;
			\rho=\suma\e\int_{\R^3}\f\,dv,\;j=\suma\e\int_{\R^3}\v\f\,dv.
			\end{gather*}
		\end{enumerate}
		\item A steady state $\left(\left(\f\right)_\alpha,\phi,A\right)$ is said to
		\begin{enumerate}[label=(\roman*)]
			\item have finite charge if
			\begin{align*}
			\int_{B_{R_0}}\int_{\R^3}\f\,dvd\left(x_1,x_2\right)<\infty
			\end{align*}
			for each $\alpha=1,\dots,N$;
			\item be compactly supported with respect to $v$ if there is $S>0$ such that $\f\left(x,v\right)=0$ for each $\alpha=1,\dots,N$, $x\in\overline\Omega$, $\left|v\right|\geq S$;
			\item be nontrivial if $\f\not\equiv 0$ for each $\alpha=1,\dots,N$;
			\item be confined with radius at most $R$ if $0<R<R_0$ such that $\f\left(x,v\right)=0$ for each $\alpha=1,\dots,N$, $x\in\overline\Omega$ with $\left|\left(x_1,x_2\right)\right|\geq R$, and $v\in\R^3$.
		\end{enumerate}
	\end{enumerate}
\end{defn}
Note that perfect conductor boundary conditions are automatically satisfied due to symmetry, as was already observed in \cref{sec:SymmConser}.
\begin{rem}
	A physically reasonable steady state should have finite charge, which usually means $\f\in L^1\left(\Omega\times\R^3\right)$ for each $\alpha=1,\dots,N$. However, this is impossible in our setting (unless all $\f$ vanish identically) by $\f$ being independent of $x_3$. Thus, here we have to modify this definition suitably as above.
\end{rem}
According to \cite{Deg90}, the natural ansatz for $\f$ is that
\begin{align}\label{eq:ansatz}
\f=\etaa\left(\E,\F,\G\right)
\end{align}
is a function of the three invariants obtained in \cref{sec:SymmConser}. We collect some basic assumptions about the ansatz functions $\etaa$:
\begin{cond}\label{cond:ansatz}
	For each $\alpha=1,\dots,N$ there holds:
	\begin{condlist}
		\item\label{cond:ansatzi} $\etaa\in C^1\left(\R^3;\left[0,\infty\right[\right)$;
		\item\label{cond:ansatzii} there exists $\etaa_\ast\in L^1\left(\R^2\right)$ such that
		\begin{align*}
		\int_{\R^2}\left|\mathcal E\etaa_\ast\left(\mathcal E, \mathcal G\right)\right|\,d\left(\mathcal E,\mathcal G\right)<\infty
		\end{align*}
		and
		\begin{align*}
		\left|\etaa\left(\mathcal E,\mathcal F,\mathcal G\right)\right|\leq\etaa_\ast\left(\mathcal E,\mathcal G\right)
		\end{align*}
		for all $\left(\mathcal E,\mathcal F,\mathcal G\right)\in\R^3$;
		\item\label{cond:ansatziii} there exists $\etaa_\#\colon\R^2\to\R$ such that
		\begin{align*}
		\forall\,d\in\R:\etaa_\#,\left|\mathcal E\right|\eta_\#\in L^1\left(\left]d,\infty\right[\times\R\right)
		\end{align*}
		and
		\begin{align*}
		\left|\nabla\etaa\left(\mathcal E,\mathcal F,\mathcal G\right)\right|\leq\etaa_\#\left(\mathcal E,\mathcal G\right)
		\end{align*}
		for all $\left(\mathcal E,\mathcal F,\mathcal G\right)\in\R^3$.
	\end{condlist}
\end{cond}
We first prove that the ansatz \cref{eq:ansatz} already ensures \cref{eq:Vlstat,eq:BCstat}. Here and in the following, we will always write $A^\tot=A+A^\ext$.
\begin{lma}\label{lma:ansatz}
	Let Conditions \ref{cond:externalpotential} and \ref{cond:ansatzi} hold and let $\phi, A_\varphi, A_3\in C^1\left(\left[0,R_0\right]\right)$ with
	\begin{align*}
	\phi'\left(0\right)=A_\varphi\left(0\right)=A_3'\left(0\right)=0.
	\end{align*}
	Then, for each $\alpha=1,\dots,N$
	\begin{align}
	\f\colon\overline\Omega\to\R,\quad\f\left(x,v\right)&=\etaa\left(\E\left(x,v\right),\F\left(x,v\right),\G\left(x,v\right)\right)\nonumber\\
	&=\etaa\left(\vo+\e\phi\left(r\right),r\left(v_\varphi+\e A_\varphi^\tot\left(r\right)\right),v_3+\e A_3^\tot\left(r\right)\right)\label{eq:ansatzex}
	\end{align}
	is continuously differentiable and satisfies \cref{eq:Vlstat,eq:BCstat}.
	\begin{proof}
		We first note that $\f$ is continuously differentiable because of $rv_\varphi=x_1v_2-x_2v_1$ and $\phi'\left(0\right)=\left(rA_\varphi^\tot\right)'\left(0\right)=\left(A_3^\tot\right)'\left(0\right)=0$. Furthermore, it is easy to see that \cref{eq:BCstat} holds since neither $\E$ nor $\F$ nor $\G$ depend on $v_r$. To ensure \cref{eq:Vlstat} for $\f$ it suffices to prove that $\E$, $\F$, and $\G$ themselves satisfy \cref{eq:Vlstat} (this clearly holds, as they are invariants of the motion; for the sake of completeness, we carry out the computation). Since they are of class $C^1$ on $\overline\Omega$, this only needs to be verified for $r>0$. In the following, have \cref{eq:EBtot} in mind. Firstly,
		\begin{align*}
		\v\cdot\partial_x\E+\e\left(E+\v\times B^\tot\right)\cdot\partial_v\E=-\e\v\cdot E+\e\left(E+\v\times B^\tot\right)\cdot\v=0.
		\end{align*}
		Secondly,
		\begin{align*}
		&\v\cdot\partial_x\F+\e\left(E+\v\times B^\tot\right)\cdot\partial_v\F\\
		&=\v\cdot\left(v_\varphi+\e A_\varphi^\tot\right)e_r-\v\cdot v_re_\varphi+\e\v\cdot r\left(A_\varphi^\tot\right)'e_r+\e\left(E+\v\times B^\tot\right)\cdot re_\varphi\\
		&=\e\v\cdot e_r\left(A_\varphi^\tot-r\left(A_\varphi^\tot\right)'+r\cdot\frac{1}{r}\left(rA_\varphi^\tot\right)'\right)=0.
		\end{align*}
		Thirdly,
		\begin{align*}
		\v\cdot\partial_x\G+\e\left(E+\v\times B^\tot\right)\cdot\partial_v\G=\e\v\cdot\left(A_3^\tot\right)'e_r+\e\left(E+\v\times B^\tot\right)\cdot e_3=0.
		\end{align*}
		Thus, \cref{eq:Vlstat} holds for $\f$ by chain rule.
	\end{proof}
\end{lma}
The ansatz \cref{eq:ansatz} in turn can be inserted into the definition of $\rho$ and $j$ to derive representations of these densities in terms of the potentials:
\begin{lma}\label{lma:reprdensities}
	Let $\phi\colon\left[0,R_0\right]\to\R$, $A\colon\left[0,R_0\right]\to\R^3$, \cref{cond:ansatzii} hold, and $\f$ be defined as in \cref{eq:ansatzex} for each $\alpha=1,\dots,N$. Then, $\f\left(x,\cdot\right)\in L^1\left(\R^3\right)$ for each $x\in\overline\Omega$. Furthermore, $\rho$ and $j$ are independent of $x_3$ and axially symmetric, and we have
	\begin{subequations}\label{eq:reprdensities}
		\begin{gather}
		4\pi\rho\left(r\right)=g_1\left(r,\phi\left(r\right),A_\varphi^\tot\left(r\right),A_3^\tot\left(r\right)\right),\\
		j_r\left(r\right)=0,\quad
		4\pi j_\varphi\left(r\right)=g_2\left(r,\phi\left(r\right),A_\varphi^\tot\left(r\right),A_3^\tot\left(r\right)\right),\quad
		4\pi j_3\left(r\right)=g_3\left(r,\phi\left(r\right),A_\varphi^\tot\left(r\right),A_3^\tot\left(r\right)\right)
		\end{gather}
	\end{subequations}
	for $r\in\left[0,R_0\right]$, where $g_1,g_2,g_3\colon\left[0,R_0\right]\times\R^3\to\R$,
	\begin{align}\label{eq:reprgdens}
	&\begin{pmatrix}g_1\\g_2\\g_3\end{pmatrix}\left(r,a,b,c\right)\nonumber\\
	&=4\pi\suma\e\int_{\R}\int_{\sqrt{\m^2+\left(\mathcal G-\e c\right)^2}+\e a}^\infty\int_0^{2\pi}\begin{pmatrix}\mathcal E-\e a\\\sqrt{\left(\mathcal E-\e a\right)^2-\left(\mathcal G-\e c\right)^2-\m^2}\sin\theta\\\mathcal G-\e c\end{pmatrix}\hspace*{3mm}\nonumber\\
	&\omit\hfill$\displaystyle\cdot\etaa\left(\mathcal E,r\sqrt{\left(\mathcal E-\e a\right)^2-\left(\mathcal G-\e c\right)^2-\m^2}\sin\theta+r\e b,\mathcal G\right)\,d\theta d\mathcal E d\mathcal G\hspace*{3mm}$\\
	&\eqqcolon\suma\,\begin{pmatrix}\g_1\\\g_2\\\g_3\end{pmatrix}\left(r,a,b,c\right)\nonumber
	\end{align}
	are continuous functions. Moreover,
	\begin{align}\label{eq:estgig1}
	\left|\left(\g_2,\g_3\right)\right|\leq\left|\g_1\right|
	\end{align}
	on $\left[0,R_0\right]\times\R^3$ for each $\alpha=1,\dots,N$.
	\begin{proof}
		At least formally we have
		\begin{align*}
		&\int_{\R^3}\begin{pmatrix}1\\\v\cdot e_r\\\v\cdot e_\varphi\\\v\cdot e_3\end{pmatrix}\,\etaa\left(\E,\F,\G\right)\,dv\\
		&=\int_{\R}\int_0^\infty\int_0^{2\pi}\frac{u}{\sqrt{\m^2+u^2+v_3^2}}\,\begin{pmatrix}\sqrt{\m^2+u^2+v_3^2}\\u\cos\theta\\u\sin\theta\\v_3\end{pmatrix}\\
		&\omit\hfill$\displaystyle\phantom{=\int_{\R}\int_0^\infty}\cdot\etaa\left(\sqrt{\m^2+u^2+v_3^2}+\e\phi\left(r\right),ru\sin\theta+r\e A_\varphi^\tot\left(r\right),v_3+\e A_3^\tot\left(r\right)\right)\,d\theta dudv_3$\\
		&=\int_{\R}\int_{\sqrt{\m^2+\left(\mathcal G-\e A_3^\tot\left(r\right)\right)^2}+\e\phi\left(r\right)}^\infty\int_0^{2\pi}\begin{pmatrix}\mathcal E-\e\phi\left(r\right)\\0\\\sqrt{\left(\mathcal E-\e\phi\left(r\right)\right)^2-\left(\mathcal G-\e A_3^\tot\left(r\right)\right)^2-\m^2}\sin\theta\\\mathcal G-\e A_3^\tot\left(r\right)\end{pmatrix}\\
		&\omit\hfill$\displaystyle\cdot\etaa\left(\mathcal E,r\sqrt{\left(\mathcal E-\e\phi\left(r\right)\right)^2-\left(\mathcal G-\e A_3^\tot\left(r\right)\right)^2-\m^2}\sin\theta+r\e A_\varphi^\tot\left(r\right),\mathcal G\right)\,d\theta d\mathcal Ed\mathcal G,$
		\end{align*}
		where we introduced polar coordinates in the $\left(v_1,v_2\right)$-plane with basis $\left(e_r,e_\varphi\right)$ and then substituted firstly $\mathcal E=\sqrt{\m^2+u^2+v_3^2}+\e\phi\left(r\right)$ and secondly $\mathcal G=v_3+\e A_3^\tot\left(r\right)$. Note that the integral in the second line vanishes after substituting $y=\sin\theta$. Due to \cref{cond:ansatzii}, the modulus of the integrand in the first line can be estimated by
		\begin{align*}
		\left(\left|\mathcal E\right|+\left|\phi\left(r\right)\right|\right)\etaa_\ast\left(\mathcal E,\mathcal G\right)
		\end{align*}
		and is hence integrable. Because of $\left|\v\right|<1$ also the other integrals exist. Thus, the above calculation is legitimated. Multiplying these identities with $\e$ and summing over $\alpha$ yields the representation. The above estimate on the integrands also implies that $g_i$ is continuous, $i=1,2,3$. Finally, \cref{eq:estgig1} is also a consequence of $\left|\v\right|<1$.
	\end{proof}
\end{lma}
\begin{rem}
	The proof of preceding lemma additionally shows that any steady state obtained in the following sections has finite charge. Indeed, for this it is sufficient that $\phi$ is integrable over $\left[0,R_0\right]$, which is of course the case when $\phi$ is continuous.
\end{rem}
In view of \cref{lma:reprdensities}, integrating \cref{eq:ODEpotentials} and using the representation \cref{eq:reprdensities}, the problem of finding a steady state with the ansatz \cref{eq:ansatz} reduces to finding $\phi$, $A_3\in C^2\left(\left[0,R_0\right]\right)$, $A_\varphi\in C^2\left(\left]0,R_0\right]\right)\cap C^1\left(\left[0,R_0\right]\right)$ satisfying \cref{eq:potentialsat0}, \cref{eq:Aphilimitr0}, and
\begin{subequations}\label{eq:inteqpotentials}
	\begin{align}
	\phi\left(r\right)&=-\int_0^r\frac{1}{s}\int_0^s\sigma g_1\left(\sigma,\phi\left(\sigma\right),A_\varphi^\tot\left(\sigma\right),A_3^\tot\left(\sigma\right)\right)\,d\sigma ds,\label{eq:inteqphi}\\
	A_\varphi\left(r\right)&=-\frac{1}{r}\int_0^rs\int_0^sg_2\left(\sigma,\phi\left(\sigma\right),A_\varphi^\tot\left(\sigma\right),A_3^\tot\left(\sigma\right)\right)\,d\sigma ds,\label{eq:inteqAphi}\\
	A_3\left(r\right)&=-\int_0^r\frac{1}{s}\int_0^s\sigma g_3\left(\sigma,\phi\left(\sigma\right),A_\varphi^\tot\left(\sigma\right),A_3^\tot\left(\sigma\right)\right)\,d\sigma ds\label{eq:inteqA3}
	\end{align}
\end{subequations}
for $r>0$ in view of \cref{lma:regularitypotentials,lma:ansatz}. Therefore, it is convenient to introduce the map
\begin{align*}
\M\colon C\left(\left[0,R_0\right];\R^3\right)&\to C\left(\left[0,R_0\right];\R^3\right),\\
\M\left(\phi,A_\varphi,A_3\right)&=\left(\left[0,R_0\right]\ni r\mapsto\begin{pmatrix}-\int_0^r\frac{1}{s}\int_0^s\sigma g_1\left(\sigma,\phi\left(\sigma\right),A_\varphi^\tot\left(\sigma\right),A_3^\tot\left(\sigma\right)\right)\,d\sigma ds\\-\frac{1}{r}\int_0^rs\int_0^sg_2\left(\sigma,\phi\left(\sigma\right),A_\varphi^\tot\left(\sigma\right),A_3^\tot\left(\sigma\right)\right)\,d\sigma ds\\-\int_0^r\frac{1}{s}\int_0^s\sigma g_3\left(\sigma,\phi\left(\sigma\right),A_\varphi^\tot\left(\sigma\right),A_3^\tot\left(\sigma\right)\right)\,d\sigma ds\end{pmatrix}\right).
\end{align*}
The following lemma shows that indeed $\M$ is well-defined (with the obvious interpretation $\M\left(\phi,A_\varphi,A_r\right)\left(0\right)=\left(0,0,0\right)$) and that it suffices to search for fixed points of $\M$:
\begin{lma}\label{lma:regpotintesol}
	Assume Conditions \ref{cond:externalpotential}, \ref{cond:ansatzi}, and \ref{cond:ansatzii}.
	\begin{lmalist}
		\item\label{lma:regpotintesoli} For any $\left(\phi,A_\varphi,A_3\right)\in C\left(\left[0,R_0\right];\R^3\right)$ we have
		\begin{align*}
		\left(\tilde\phi,\tilde A_\varphi,\tilde A_3\right)\coloneqq\M\left(\phi,A_\varphi,A_3\right)\in C^2\left(\left[0,R_0\right];\R^3\right).
		\end{align*} 
		Furthermore, $\left(\tilde\phi,\tilde A_\varphi,\tilde A_3\right)$ satisfies \cref{eq:potentialsat0,eq:Aphilimitr0}.
		\item If $\left(\phi,A_\varphi,A_3\right)\in C\left(\left[0,R_0\right];\R^3\right)$ is a fixed point of $\M$, then $\left(\left(\f\right)_\alpha,\phi,A\right)$ is a steady state, where the $\f$ are defined via the ansatz \cref{eq:ansatz}.
	\end{lmalist}
	\begin{proof}
		Due to \cref{lma:reprdensities}, the functions 
		\begin{align*}
		\tilde g_i\colon\left[0,R_0\right]\to\R,\quad\tilde g_i\left(\sigma\right)=g_i\left(\sigma,\phi\left(\sigma\right),A_\varphi^\tot\left(\sigma\right),A_3^\tot\left(\sigma\right)\right)
		\end{align*}
		are continuous, $i=1,2,3$, and hence bounded by some constant $C>0$. Thus, there holds
		\begin{align*}
		\left|\tilde\phi\left(r\right)\right|,\left|\tilde A_3\left(r\right)\right|\leq C\int_0^r\frac{1}{s}\int_0^s\sigma d\sigma=\frac{C}{4}r^2,\quad\left|\tilde A_\varphi\left(r\right)\right|\leq\frac{C}{r}\int_0^rs\int_0^sd\sigma=\frac{C}{3}r^2.
		\end{align*}
		Hence, $\tilde\phi$, $\tilde A_\varphi$, and $\tilde A_3$ are continuous also at $r=0$, and $\frac{\tilde A_\varphi\left(r\right)}{r}=\mathcal O\left(r\right)$ for $r\to 0$. Furthermore, the \enquote*{tilde}-potentials are twice continuously differentiable on $\left]0,R_0\right[$ with
		\begin{align*}
		\tilde\phi'\left(r\right)&=-\frac{1}{r}\int_0^rs\tilde g_1\left(s\right)\,ds,\quad\tilde\phi''\left(r\right)=\frac{1}{r^2}\int_0^rs\tilde g_1\left(s\right)\,ds-\tilde g_1\left(r\right),\\
		\tilde A_\varphi'\left(r\right)&=\frac{1}{r^2}\int_0^rs\int_0^s\tilde g_2\left(\sigma\right)\,d\sigma ds-\int_0^r\tilde g_2\left(s\right)\,ds,\\
		\tilde A_\varphi''\left(r\right)&=-\frac{2}{r^3}\int_0^rs\int_0^s\tilde g_2\left(\sigma\right)\,d\sigma ds+\frac{1}{r}\int_0^r\tilde g_2\left(s\right)\,ds-\tilde g_2\left(r\right),\\
		\tilde A_3'\left(r\right)&=-\frac{1}{r}\int_0^rs\tilde g_3\left(s\right)\,ds,\quad\tilde A_3''\left(r\right)=\frac{1}{r^2}\int_0^rs\tilde g_3\left(s\right)\,ds-\tilde g_3\left(r\right).
		\end{align*}
		Because of
		\begin{align*}
		\left|\tilde\phi'\left(r\right)\right|,\left|\tilde A_3'\left(r\right)\right|\leq\frac{C}{r}\int_0^rs\,ds=\frac{C}{2}r,\quad\left|\tilde A_\varphi'\left(r\right)\right|\leq\frac{C}{r^2}\int_0^rs\int_0^sd\sigma+Cr=\frac{4C}{3}r
		\end{align*}
		they are continuously differentiable on whole $\left[0,R_0\right]$ with vanishing derivative at $r=0$, and moreover $\tilde A_\varphi'\left(r\right)=\mathcal O\left(r\right)$ for $r\to 0$. Furthermore, by l'Hôpital's rule we have
		\begin{align*}
		\lim_{r\to 0}\tilde\phi''\left(r\right)&=\lim_{r\to 0}\frac{r\tilde g_1\left(r\right)}{2r}-\tilde g_1\left(0\right)=-\frac{\tilde g_1\left(0\right)}{2},\\
		\lim_{r\to 0}\tilde A_\varphi''\left(r\right)&=-\lim_{r\to 0}\frac{2r\int_0^r\tilde g_2\left(s\right)\,ds}{3r^2}+\tilde g_2\left(0\right)-\tilde g_2\left(0\right)=-\frac{2\tilde g_2\left(0\right)}{3},\\
		\lim_{r\to 0}\tilde A_3''\left(r\right)&=\lim_{r\to 0}\frac{r\tilde g_3\left(r\right)}{2r}-\tilde g_3\left(0\right)=-\frac{\tilde g_3\left(0\right)}{2}.
		\end{align*}
		Therefore, $\tilde\phi,\tilde A_\varphi,\tilde A_3\in C^2\left(\left[0,R_0\right]\right)$ and clearly $\tilde A_\varphi''\left(r\right)=\mathcal O\left(1\right)$ for $r\to 0$. Finally, from \cref{lma:regularitypotentials,lma:ansatz,lma:reprdensities} follows that $\left(\left(\f\right)_\alpha,\phi,A\right)$ is a steady state if $\left(\phi,A_\varphi,A_3\right)$ is a fixed point of $\M$; note that \cref{eq:inteqpotentials} implies \cref{eq:ODEpotentials} and this yields $-\Delta_x\phi=4\pi\rho$ on $\overline\Omega$ and $-\Delta_xA=4\pi j$ on $\overline\Omega\setminus\R e_3$ in the classical sense, and $-\Delta_xA=4\pi j$ on $\overline\Omega$ in the weak sense.
	\end{proof}
\end{lma}

\section{Existence of steady states}\label{sec:existence}
\subsection{A priori estimates}\label{sec:apriori}
Hence, there only remains to find a fixed point of $\M$. For this, the most important tool is to derive a priori bounds for the potentials. Therefore, we assume that we already have a solution $\left(\phi,A_\varphi,A_3\right)\in C\left(\left[0,R_0\right];\R^3\right)$ of \cref{eq:inteqpotentials} for the time being. Due to \cref{eq:reprgdens}, we first have the following estimate on $\g_1$ for each $\left(r,a,b,c\right)\in\left[0,R_0\right]\times\R^3$:
\begin{align*}
\left|\g_1\left(r,a,b,c\right)\right|\leq 4\pi\left|\e\right|\cdot 2\pi\int_{\R^2}\left(\left|\mathcal E\right|+\left|\e\right|\left|a\right|\right)\etaa_\ast\left(\mathcal E,\mathcal G\right)\,d\left(\mathcal E,\mathcal G\right).
\end{align*}
Using \cref{eq:estgig1} and summing over $\alpha$ yield
\begin{align}\label{eq:estgi}
\left|g_i\left(r,a,b,c\right)\right|\leq c_1+c_2\left|a\right|,\quad i=1,2,3,
\end{align}
where we introduced the abbreviations
\begin{align*}
c_1\coloneqq 8\pi^2\suma\left|\e\right|\int_{\R^2}\left|\mathcal E\right|\etaa_\ast\left(\mathcal E,\mathcal G\right)\,d\left(\mathcal E,\mathcal G\right)<\infty,\quad
c_2\coloneqq 8\pi^2\suma\left|\e\right|^2\int_{\R^2}\etaa_\ast\left(\mathcal E,\mathcal G\right)\,d\left(\mathcal E,\mathcal G\right)<\infty.
\end{align*}
Therefore, in view of \cref{eq:inteqphi} an integral inequality for $\phi$ follows, in particular
\begin{align}\label{eq:intineqphi}
\left|\phi\left(r\right)\right|\leq\int_0^r\frac{1}{s}\int_0^s\sigma\left(c_1+c_2\left|\phi\left(\sigma\right)\right|\right)\,d\sigma ds=\frac{c_1}{4}r^2+c_2\int_0^r\frac{1}{s}\int_0^s\sigma\left|\phi\left(\sigma\right)\right|\,d\sigma ds
\end{align}
for $r\in\left[0,R_0\right]$. We could thus easily derive the inequality
\begin{align}\label{eq:estphiintfalse}
\left|\phi\left(r\right)\right|\leq\frac{c_1}{4}R_0^2+c_2R_0\int_0^r\left|\phi\left(s\right)\right|\,ds
\end{align}
and therefore
\begin{align}\label{eq:estphifalse}
\left|\phi\left(r\right)\right|\leq\frac{c_1}{4}R_0^2e^{c_2R_0r}
\end{align}
via Gronwall. However, \cref{eq:estphiintfalse} is way too crude and hence \cref{eq:estphifalse} is not very sharp. If we were to use this a priori estimate later to show confinement of a steady state, the needed assumption about the external potential would be quite strong. Consequently, in order to allow a wider class for external potentials ensuring confinement later, we now search for a sharper a priori estimate on $\phi$.

Thus, we search for a solution of the integral equation corresponding to \cref{eq:intineqphi}, that is,
\begin{align}\label{eq:inteqxi}
\xi\left(r\right)=\frac{c_1}{4}r^2+c_2\int_0^r\frac{1}{s}\int_0^s\sigma\xi\left(\sigma\right)\,d\sigma ds.
\end{align}
For any $\xi\in C\left(\left[0,R_0\right]\right)$, there holds the elementary identity
\begin{align}\label{eq:elementaryidentity}
\int_0^r\frac{1}{s}\int_0^s\sigma\xi\left(\sigma\right)\,d\sigma ds=\int_0^r\left(\ln r-\ln s\right)s\xi\left(s\right)\,ds
\end{align}
for any $\left[0,R_0\right]$, which can easily be verified by differentiating both sides with respect to $r$ and noting that both sides vanish for $r=0$. Therefore, \cref{eq:inteqxi} becomes an Volterra integral equation of the second kind, in particular
\begin{align}\label{eq:inteqxiVolterra}
\xi\left(r\right)=\frac{c_1}{4}r^2+c_2\int_0^r\left(\ln r-\ln s\right)s\xi\left(s\right)\,ds
\end{align}
with nonnegative, square integrable Volterra kernel
\begin{align*}
V\colon\left[0,R_0\right]^2\to\R,\quad V\left(r,s\right)=\begin{cases}c_2\left(\ln r-\ln s\right)s,&0<s\leq r\leq R_0,\\0,&\text{else}.\end{cases}
\end{align*}
It is well known that Volterra integral equations such as \cref{eq:inteqxiVolterra} have a unique square integrable solution, see \cite[Sec. 1.5.]{Tri85}. To find this solution, we rather work with \cref{eq:inteqxi}, which suggests a series ansatz
\begin{align*}
\xi\left(r\right)=\sum_{k=0}^\infty a_kr^k
\end{align*}
for $\xi$. With this ansatz, at least formally we demand
\begin{align}
\sum_{k=0}^\infty a_kr^k&\overset{!}=\frac{c_1}{4}r^2+c_2\int_0^r\frac{1}{s}\int_0^s\sigma\xi\left(\sigma\right)\,d\sigma ds=\frac{c_1}{4}r^2+c_2\int_0^r\frac{1}{s}\int_0^s\sigma\sum_{k=0}^\infty a_k\sigma^k\,d\sigma ds\nonumber\\
&=\frac{c_1}{4}r^2+c_2\int_0^r\sum_{k=0}^\infty\frac{a_k}{k+2}s^{k+1}\,ds=\frac{c_1}{4}r^2+c_2\int_0^r\sum_{k=0}^\infty\frac{a_k}{k+2}s^{k+1}\,ds\nonumber\\
&=\frac{c_1}{4}r^2+c_2\sum_{k=0}^\infty\frac{a_k}{(k+2)^2}r^{k+2}=\frac{c_1}{4}r^2+\sum_{k=2}^\infty\frac{c_2a_{k-2}}{k^2}r^k.\label{eq:xiansatzeq}
\end{align}
Thus,
\begin{align*}
a_0=a_1=0,\quad a_2=\frac{c_1}{4}+\frac{c_2a_0}{2^2}=\frac{c_1}{4}.
\end{align*}
Therefore, $a_k=0$ if $k$ is odd, and
\begin{align*}
a_{2m}=\frac{c_2a_{2\left(m-1\right)}}{4m^2}
\end{align*}
for $m\geq 2$. Hence, we have
\begin{align*}
a_{2m}=\frac{c_1c_2^{m-1}}{4^m\left(m!\right)^2}
\end{align*}
for $m\in\N$ by induction. Consequently, we define
\begin{align*}
\xi\colon\R\to\R,\quad \xi\left(r\right)=\sum_{k=1}^\infty\frac{c_1c_2^{k-1}}{4^k\left(k!\right)^2}r^{2k}.
\end{align*}
Obviously, this series is uniformly convergent on any bounded interval, whence the calculation \cref{eq:xiansatzeq} is legitimated and $\xi$ indeed is the unique square integrable solution of \cref{eq:inteqxiVolterra} on $\left[0,R_0\right]$ by \cref{eq:elementaryidentity}. Moreover, $\phi$ satisfies the corresponding integral inequality
\begin{align*}
\left|\phi\left(r\right)\right|\leq\frac{c_1}{4}r^2+c_2\int_0^r\left(\ln r-\ln s\right)s\left|\phi\left(s\right)\right|\,ds.
\end{align*}
Thus, there holds
\begin{align}\label{eq:phiapriori}
\left|\phi\left(r\right)\right|\leq\xi\left(r\right)
\end{align}
for all $r\in\left[0,R_0\right]$ as a consequence of the positivity of Volterra operators in the case $V\geq 0$, cf. \cite[Theorem 5]{Bee69}. Therefore, we have established a quite sharp a priori bound on $\phi$.

In order to obtain similar estimates also for $A_\varphi$ and $A_3$, we insert \cref{eq:estgi,eq:phiapriori} into \cref{eq:inteqAphi,eq:inteqA3}. On the one hand, we conclude
\begin{align}\label{eq:Aphiapriori}
\left|A_\varphi\left(r\right)\right|&\leq\frac{1}{r}\int_0^rs\int_0^s\left(c_1+c_2\left|\phi\left(\sigma\right)\right|\right)\,d\sigma ds\leq\frac{c_1}{3}r^2+\frac{c_2}{r}\int_0^rs\int_0^s\xi\left(\sigma\right)\,d\sigma ds\nonumber\\
&=\frac{c_1}{3}r^2+\frac{c_2}{r}\int_0^r\sum_{k=1}^\infty\frac{c_1c_2^{k-1}}{\left(2k+1\right)4^k\left(k!\right)^2}s^{2k+2}\,ds\nonumber\\
&=\frac{c_1}{3}r^2+\sum_{k=1}^\infty\frac{c_1c_2^k}{\left(2k+1\right)\left(2k+3\right)4^k\left(k!\right)^2}r^{2k+2}=\sum_{k=1}^\infty\frac{c_1c_2^{k-1}}{\left(1-\frac{1}{4k^2}\right)4^k\left(k!\right)^2}r^{2k}\eqqcolon\zeta\left(r\right)
\end{align}
and on the other hand
\begin{align}\label{eq:A3apriori}
\left|A_3\left(r\right)\right|\leq\int_0^r\frac{1}{s}\int_0^s\sigma\left(c_1+c_2\left|\phi\left(\sigma\right)\right|\right)\,d\sigma ds\leq\frac{c_1}{4}r^2+c_2\int_0^r\frac{1}{s}\int_0^s\sigma\xi\left(\sigma\right)\,d\sigma ds=\xi\left(r\right)
\end{align}
for $r\in\left[0,R_0\right]$. Note that the a priori bound on $A_\varphi$ is slightly weaker than the bounds on $\phi$ and $A_3$, since obviously $\xi\leq\zeta$.

Thus, we have proved the following important a priori estimate:
\begin{lma}\label{lma:apriori}
	Let $\left(\phi,A_\varphi,A_3\right)\in C\left(\left[0,R_0\right];\R^3\right)$ be a fixed point of $\M$. Then there holds
	\begin{align*}
	\left|\phi\left(r\right)\right|,\left|A_3\left(r\right)\right|\leq\xi\left(r\right),\quad\left|A_\varphi\left(r\right)\right|\leq\zeta\left(r\right)
	\end{align*}
	for $r\in\left[0,R_0\right]$.
\end{lma}

For the sake of completeness, we remark that $\xi$ can be written in terms of a Bessel function, which corresponds to the fact that \cref{eq:inteqxi} implies
\begin{align*}
r^2\xi''+r\xi'-c_2r^2\xi=c_1r^2,
\end{align*}
whence
\begin{align*}
z\left(r\right)\coloneqq\frac{c_2}{c_1}\xi\left(\frac{r}{\sqrt{c_2}}\right)+1
\end{align*}
solves the modified Bessel equation
\begin{align*}
r^2z''+rz'-r^2z=0.
\end{align*}
Endowed with the initial condition $\xi\left(0\right)=\xi'\left(0\right)=0$, this yields $z=I_0$, where $I_0$ is the modified Bessel function of the first kind (with parameter $0$). Consequently,
\begin{align*}
\xi\left(r\right)=\frac{c_1}{c_2}\left(I_0\left(\sqrt{c_2}r\right)-1\right).
\end{align*}

\subsection{Fixed point argument}
We proceed with proving that steady states really do exist via some fixed point argument. Throughout the rest of this section, we assume that \cref{cond:ansatz} holds and equip the space $C\left(\left[0,R_0\right];\R^3\right)$ with the norm
\begin{align*}
\left\|\left(\phi,A_\varphi,A_3\right)\right\|_{C\left(\left[0,R_0\right];\R^3\right)}=\left(\left\|\phi\right\|_{C\left(\left[0,R_0\right]\right)}^2+\left\|A_\varphi\right\|_{C\left(\left[0,R_0\right]\right)}^2+\left\|A_3\right\|_{C\left(\left[0,R_0\right]\right)}^2\right)^{\frac{1}{2}}.
\end{align*}
The a priori bounds obtained in the last section are an important tool to prove existence of solutions to \cref{eq:inteqpotentials}. In view of Schaefer's fixed point theorem (see \cite[Sec. 9.2.2.]{Eva10} for example), we have to prove that $\M$ is continuous and compact, and we have to establish a priori bounds on possible fixed points of the operators $\lambda\M$ for $0\leq\lambda\leq 1$. The second task is easily carried out by using the results of \cref{sec:apriori}:
\begin{lma}\label{lma:apriorilambda}
	Let $\left(\phi,A_\varphi,A_3\right)\in C\left(\left[0,R_0\right];\R^3\right)$ such that $\left(\phi,A_\varphi,A_3\right)=\lambda\M\left(\phi,A_\varphi,A_3\right)$ for some $0\leq\lambda\leq 1$. Then there holds
	\begin{align*}
	\left|\phi\left(r\right)\right|,\left|A_3\left(r\right)\right|\leq\xi\left(r\right),\quad\left|A_\varphi\left(r\right)\right|\leq\zeta\left(r\right)
	\end{align*}
	for $r\in\left[0,R_0\right]$. In particular, the set 
	\begin{align*}
	\left\{\left(\phi,A_\varphi,A_3\right)\in C\left(\left[0,R_0\right];\R^3\right)\mid\left(\phi,A_\varphi,A_3\right)=\lambda\M\left(\phi,A_\varphi,A_3\right)\text{ for some }0\leq\lambda\leq 1\right\}
	\end{align*}
	is bounded.
	\begin{proof}
		By \cref{eq:estgi}, we obtain
		\begin{align*}
		\left|\phi\left(r\right)\right|\leq\lambda\int_0^r\frac{1}{s}\int_0^s\sigma\left(c_1+c_2\left|\phi\left(\sigma\right)\right|\right)\,d\sigma ds\leq\frac{c_1}{4}r^2+c_2\int_0^r\frac{1}{s}\int_0^s\sigma\left|\phi\left(\sigma\right)\right|\,d\sigma ds
		\end{align*}
		similarly to \cref{eq:intineqphi}. Hence, there holds $\left|\phi\left(r\right)\right|\leq\xi\left(r\right)$ for $r\in\left[0,R_0\right]$. Similarly to \cref{eq:Aphiapriori,eq:A3apriori}, we also have
		\begin{align*}
		\left|A_\varphi\left(r\right)\right|&\leq\frac{\lambda}{r}\int_0^rs\int_0^s\left(c_1+c_2\left|\phi\left(\sigma\right)\right|\right)\,d\sigma ds\leq\frac{c_1}{3}r^2+\frac{c_2}{r}\int_0^rs\int_0^s\xi\left(\sigma\right)\,d\sigma ds=\zeta\left(r\right),\\
		\left|A_3\left(r\right)\right|&\leq\lambda\int_0^r\frac{1}{s}\int_0^s\sigma\left(c_1+c_2\left|\phi\left(\sigma\right)\right|\right)\,d\sigma ds\leq\frac{c_1}{4}r^2+c_2\int_0^r\frac{1}{s}\int_0^s\sigma\xi\left(\sigma\right)\,d\sigma ds=\xi\left(r\right)
		\end{align*}
		for $r\in\left[0,R_0\right]$.
	\end{proof}
\end{lma}
Thus, there remains to prove the following lemma:
\begin{lma}\label{lma:Mcontcomp}
	The map $\M$ is (even locally Lipschitz) continuous and compact.
	\begin{proof}
		Let $S>0$ and $\left(\phi,A_\varphi,A_3\right),\left(\overline\phi,\overline A_\varphi,\overline A_3\right)\in\overline{B_S}\subset C\left(\left[0,R_0\right];\R^3\right)$ (here and in the following, $\overline{B_S}$ denotes the closed ball about the origin with radius $S$). On the one hand, following the calculation in the proof of \cref{lma:reprdensities}, we have for each $r\in\left[0,R_0\right]$ for some $\left(a,b,c\right)$ in the line segment connecting $\left(\phi\left(r\right),A_\varphi\left(r\right),A_3\left(r\right)\right)$ and $\left(\overline\phi\left(r\right),\overline A_\varphi\left(r\right),\overline A_3\left(r\right)\right)$,
		\begin{align}
		&\left|\left(g_1,g_2,g_3\right)\left(r,\phi\left(r\right),A_\varphi^\tot\left(r\right),A_3^\tot\left(r\right)\right)-\left(g_1,g_2,g_3\right)\left(r,\overline\phi\left(r\right),\overline A_\varphi^\tot\left(r\right),\overline A_3^\tot\left(r\right)\right)\right|\nonumber\\
		&=\left|4\pi\suma\e^2\int_\R\int_0^\infty\int_0^{2\pi}\frac{u}{\sqrt{\m^2+u^2+v_3^2}}\begin{pmatrix}\sqrt{\m^2+u^2+v_3^2}\\u\sin\theta\\v_3\end{pmatrix}\right.\nonumber\\
		&\phantom{=\quad}\cdot\left[\vphantom{\begin{pmatrix}\phi\left(r\right)-\overline\phi\left(r\right)\\r\left(A_\varphi\left(r\right)-\overline A_\varphi\left(r\right)\right)\\A_3\left(r\right)-\overline A_3\left(r\right)\end{pmatrix}}\nabla\etaa\left(\sqrt{\m^2+u^2+v_3^2}+\e a,ru\sin\theta+r\e b+r\e A_\varphi^\ext\left(r\right),v_3+\e c+\e A_3^\ext\left(r\right)\right)\right.\nonumber\\
		&\omit\hfill$\displaystyle\left.\left.\cdot\begin{pmatrix}\phi\left(r\right)-\overline\phi\left(r\right)\\r\left(A_\varphi\left(r\right)-\overline A_\varphi\left(r\right)\right)\\A_3\left(r\right)-\overline A_3\left(r\right)\end{pmatrix}\right]\,d\theta dudv_3\right|$\nonumber\\
		&=\left|\vphantom{\begin{pmatrix}\phi\left(r\right)-\overline\phi\left(r\right)\\r\left(A_\varphi\left(r\right)-\overline A_\varphi\left(r\right)\right)\\A_3\left(r\right)-\overline A_3\left(r\right)\end{pmatrix}}4\pi\suma\e^2\int_\R\int_{\sqrt{\m^2+\left(\mathcal G-\e c-\e A_3^\ext\left(r\right)\right)^2}+\e a}^\infty\int_0^{2\pi}\right.\nonumber\\
		&\omit\hfill$\displaystyle\begin{pmatrix}\mathcal E-\e a\\\sqrt{\left(\mathcal E-\e a\right)^2-\left(\mathcal G-\e c-\e A_3^\ext\left(r\right)\right)^2-\m^2}\sin\theta\\\mathcal G-\e c-\e A_3^\ext\left(r\right)\end{pmatrix}$\nonumber\\
		&\phantom{\;}\cdot\left[\vphantom{\begin{pmatrix}\phi\left(r\right)-\overline\phi\left(r\right)\\r\left(A_\varphi\left(r\right)-\overline A_\varphi\left(r\right)\right)\\A_3\left(r\right)-\overline A_3\left(r\right)\end{pmatrix}}\nabla\etaa\left(\mathcal E,r\sqrt{\left(\mathcal E-\e a\right)^2-\left(\mathcal G-\e c-\e A_3^\ext\left(r\right)\right)^2-\m^2}\sin\theta+r\e b+r\e A_\varphi^\ext\left(r\right),\mathcal G\right)\right.\nonumber\\
		&\omit\hfill$\displaystyle\left.\left.\cdot\begin{pmatrix}\phi\left(r\right)-\overline\phi\left(r\right)\\r\left(A_\varphi\left(r\right)-\overline A_\varphi\left(r\right)\right)\\A_3\left(r\right)-\overline A_3\left(r\right)\end{pmatrix}\right]\,d\theta d\mathcal Ed\mathcal G\right|$\nonumber\\
		&\leq 8\sqrt{3}\pi^2\left(1+R_0\right)\suma\left|\e\right|^2\int_\R\int_{-S}^\infty\left(\left|\mathcal E\right|+S\right)\etaa_\#\left(\mathcal E,\mathcal G\right)\,d\mathcal Ed\mathcal G\cdot\left|\left(\phi,A_\varphi,A_3\right)\left(r\right)-\left(\overline\phi,\overline A_\varphi,\overline A_3\right)\left(r\right)\right|\nonumber\\
		&=C\left(S\right)\left|\left(\phi,A_\varphi,A_3\right)\left(r\right)-\left(\overline\phi,\overline A_\varphi,\overline A_3\right)\left(r\right)\right|,\label{eq:gLipschitz}
		\end{align}
		where the constant $C\left(S\right)$ is finite due to \cref{cond:ansatziii}. Integrating this estimate, we conclude
		\begin{align*}
		&\left|\M\left(\phi,A_\varphi,A_3\right)\left(r\right)-\M\left(\overline\phi,\overline A_\varphi,\overline A_3\right)\left(r\right)\right|\\
		&\leq C\left(S\right)\left\|\left(\phi,A_\varphi,A_3\right)-\left(\overline\phi,\overline A_\varphi,\overline A_3\right)\right\|_{C\left(\left[0,R_0\right];\R^3\right)}\\
		&\phantom{\leq C\left(S\right)\;}\cdot\left|\left(\int_0^r\frac{1}{s}\int_0^s\sigma\,d\sigma ds,\frac{1}{r}\int_0^rs\int_0^sd\sigma ds,\int_0^r\frac{1}{s}\int_0^s\sigma\,d\sigma ds\right)\right|\\
		&=C\left(S\right)\cdot\frac{\sqrt{34}}{12}r^2\left\|\left(\phi,A_\varphi,A_3\right)-\left(\overline\phi,\overline A_\varphi,\overline A_3\right)\right\|_{C\left(\left[0,R_0\right];\R^3\right)},
		\end{align*}
		whence
		\begin{align*}
		&\left\|\M\left(\phi,A_\varphi,A_3\right)-\M\left(\overline\phi,\overline A_\varphi,\overline A_3\right)\right\|_{C\left(\left[0,R_0\right];\R^3\right)}\\
		&\leq C\left(S\right)\cdot\frac{\sqrt{34}}{12}R_0^2\left\|\left(\phi,A_\varphi,A_3\right)-\left(\overline\phi,\overline A_\varphi,\overline A_3\right)\right\|_{C\left(\left[0,R_0\right];\R^3\right)}.
		\end{align*}
		Therefore, $\M$ is locally Lipschitz continuous.
		
		On the other hand, by \cref{eq:estgi} we have
		\begin{align*}
		\left|g_i\left(r,\phi\left(r\right),A_\varphi^\tot\left(r\right),A_3^\tot\left(r\right)\right)\right|\leq c_1+c_2\left|\phi\left(r\right)\right|\leq c_1+c_2S\eqqcolon\tilde C\left(S\right)
		\end{align*}
		for $i=1,2,3$ and $r\in\left[0,R_0\right]$. Furthermore, there holds
		\begin{align*}
		\left(\M\left(\phi,A_\varphi,A_3\right)\right)'\left(0\right)=\left(0,0,0\right)
		\end{align*}
		by (the proof of) \cref{lma:regpotintesoli} and for $0<r\leq R_0$
		\begin{align*}
		\left|\left(\M_i\left(\phi,A_\varphi,A_3\right)\right)'\left(r\right)\right|=\left|-\frac{1}{r}\int_0^rsg_i\left(s,\phi\left(s\right),A_\varphi^\tot\left(s\right),A_3^\tot\left(s\right)\right)\,ds\right|\leq\frac{\tilde C\left(S\right)r}{2}\leq\frac{\tilde C\left(S\right)R_0}{2}
		\end{align*}
		for $i=1,3$ and
		\begin{align*}
		&\left|\left(\M_2\left(\phi,A_\varphi,A_3\right)\right)'\left(r\right)\right|\\
		&=\left|\frac{1}{r^2}\int_0^rs\int_0^sg_2\left(\sigma,\phi\left(\sigma\right),A_\varphi^\tot\left(\sigma\right),A_3^\tot\left(\sigma\right)\right)\,d\sigma ds-\int_0^rg_2\left(s,\phi\left(s\right),A_\varphi^\tot\left(s\right),A_3^\tot\left(s\right)\right)\,ds\right|\\
		&\leq\frac{\tilde C\left(S\right)r}{3}+\tilde C\left(S\right)r\leq\frac{4\tilde C\left(S\right)R_0}{3}.
		\end{align*}
		Therefore, for each $\left(\phi,A_\varphi,A_3\right)\in\overline{B_S}$, we have that $\mathcal M\left(\phi,A_\varphi,A_3\right)$ is Lipschitz continuous with a uniform Lipschitz constant, i.e., a Lipschitz constant only depending on $S$. By the theorem of Arzelà-Ascoli, $\M$ thus maps bounded sets to precompact sets, that is, $\M$ is compact.
	\end{proof}
\end{lma}
\begin{thm}\label{thm:MfixedpointSteadyState}
	Let \cref{cond:externalpotential,cond:ansatz} hold. Then, $\M$ has a unique fixed point. Thus, there exists an axially symmetric steady state $\left(\left(\f\right)_\alpha,\phi,A\right)$ of the two and one-half dimensional relativistic Vlasov-Maxwell system on $\overline\Omega$ with external potential $A^\ext$, where the $\f$ are written in terms of $\phi$ and $A$, cf. \cref{eq:ansatzex}.
	\begin{proof}
		Combining \cref{lma:apriorilambda,lma:Mcontcomp} and invoking Schaefer's fixed point theorem we conclude that $\M$ has a fixed point. Due to \cref{lma:regpotintesol}, we obtain a corresponding steady state.
		
		There remains to prove that a fixed point of $\M$ is unique. If we have two fixed points $\left(\phi,A_\varphi,A_3\right)$, $\left(\overline\phi,\overline A_\varphi,\overline A_3\right)$ of $\M$, let $S>0$ such that 
		\begin{align*}
			\left(\phi,A_\varphi,A_3\right),\left(\overline\phi,\overline A_\varphi,\overline A_3\right)\in\overline{B_S}\subset C\left(\left[0,R_0\right];\R^3\right).
		\end{align*}
		By \cref{eq:gLipschitz} and $0\leq\sigma\leq s\leq r\leq R_0$ there holds
		\begin{align*}
		&\left|\left(\phi,A_\varphi,A_3\right)\left(r\right)-\left(\overline\phi,\overline A_\varphi,\overline A_3\right)\left(r\right)\right|=\left|\left(\M\left(\phi,A_\varphi,A_3\right)\right)\left(r\right)-\left(\M\left(\overline\phi,\overline A_\varphi,\overline A_3\right)\right)\left(r\right)\right|\\
		&\leq C\left(S\right)\left|\left(\int_0^r\frac{1}{s}\int_0^s\sigma\left|\left(\phi,A_\varphi,A_3\right)\left(\sigma\right)-\left(\overline\phi,\overline A_\varphi,\overline A_3\right)\left(\sigma\right)\right|\,d\sigma ds,\right.\right.\\
		&\phantom{\leq\; C\left(S\right)\Bigg|\Bigg(}\frac{1}{r}\int_0^rs\int_0^s\left|\left(\phi,A_\varphi,A_3\right)\left(\sigma\right)-\left(\overline\phi,\overline A_\varphi,\overline A_3\right)\left(\sigma\right)\right|d\sigma ds,\\
		&\phantom{\leq\; C\left(S\right)\Bigg|\Bigg(}\left.\left.\int_0^r\frac{1}{s}\int_0^s\sigma\left|\left(\phi,A_\varphi,A_3\right)\left(\sigma\right)-\left(\overline\phi,\overline A_\varphi,\overline A_3\right)\left(\sigma\right)\right|\,d\sigma ds\right)\right|\\
		&\leq C\left(S\right)\cdot\sqrt{3}R_0\int_0^r\left|\left(\phi,A_\varphi,A_3\right)\left(s\right)-\left(\overline\phi,\overline A_\varphi,\overline A_3\right)\left(s\right)\right|\,ds
		\end{align*}
		for each $r\in\left[0,R_o\right]$. Thus, the two fixed points coincide due to Gronwall's lemma.
	\end{proof}
\end{thm}

\subsection{Direct construction}
Since the above proof of existence of steady states is not constructive, we now provide a method to obtain steady states which is constructive. To this end, we define an approximating sequence $\left(\left(\phi^k,A_\varphi^k,A_3^k\right)\right)_{k\in\N_0}$ recursively via
\begin{align*}
\left(\phi^0,A_\varphi^0,A_3^0\right)=\left(0,0,0\right),\quad\left(\phi^{k+1},A_\varphi^{k+1},A_3^{k+1}\right)=\M\left(\phi^k,A_\varphi^k,A_3^k\right).
\end{align*}
To show that this sequence indeed converges to a (and thus the) fixed point of $\M$, we first prove that this sequence is bounded. In fact, the a priori estimates of \cref{sec:apriori} carry over:
\begin{lma}\label{lma:aprioriiterates}
	For each $k\in\N_0$ and $r\in\left[0,R_0\right]$ there holds
	\begin{align*}
	\left|\phi^k\left(r\right)\right|,\left|A_3^k\left(r\right)\right|\leq\xi\left(r\right),\quad\left|A_\varphi^k\left(r\right)\right|\leq\zeta\left(r\right).
	\end{align*}
	In particular,
	\begin{align*}
	\left\|\left(\phi^k,A_\varphi^k,A_3^k\right)\right\|_{C\left(\left[0,R_0\right];\R^3\right)}\leq\sqrt{2\xi\left(R_0\right)^2+\zeta\left(R_0\right)^2}\eqqcolon S.
	\end{align*}
	\begin{proof}
		We prove
		\begin{align*}
		\left|\phi^k\left(r\right)\right|,\left|A_3^k\left(r\right)\right|\leq\sum_{j=1}^k\frac{c_1c_2^{j-1}}{4^j\left(j!\right)^2}r^{2j},\quad\left|A_\varphi^k\left(r\right)\right|\leq\sum_{j=1}^k\frac{c_1c_2^{j-1}}{\left(1-\frac{1}{4j^2}\right)4^j\left(j!\right)^2}r^{2j}
		\end{align*}
		via induction, from which the assertion follows. Indeed, this obviously holds true for $k=0$, and thanks to \cref{eq:estgi} we also have
		\begin{align*}
		&\left|\phi^{k+1}\left(r\right)\right|,\left|A_3^{k+1}\left(r\right)\right|\leq\int_0^r\frac{1}{s}\int_0^s\sigma\left(c_1+c_2\left|\phi^k\left(\sigma\right)\right|\right)\,d\sigma ds\\
		&\leq\frac{c_1}{4}r^2+c_2\int_0^r\frac{1}{s}\int_0^s\sigma\sum_{j=1}^k\frac{c_1c_2^{j-1}}{4^j\left(j!\right)^2}\sigma^{2j}\,d\sigma ds=\frac{c_1}{4}r^2+c_2\int_0^r\sum_{j=1}^k\frac{c_1c_2^{j-1}}{4^j\left(j!\right)^2\left(2j+2\right)}s^{2j+1}\,ds\\
		&=\frac{c_1}{4}r^2+\sum_{j=1}^k\frac{c_1c_2^j}{4^{j+1}\left(\left(j+1\right)!\right)^2}r^{2j+2}=\sum_{j=1}^{k+1}\frac{c_1c_2^{j-1}}{4^j\left(j!\right)^2}r^{2j}
		\end{align*}
		and
		\begin{align*}
		\left|A_\varphi^{k+1}\left(r\right)\right|&\leq\frac{1}{r}\int_0^rs\int_0^\sigma\left(c_1+c_2\left|\phi^k\left(\sigma\right)\right|\right)\,d\sigma ds\leq\frac{c_1}{3}r^2+\frac{c_2}{r}\int_0^rs\int_0^s\sum_{j=1}^k\frac{c_1c_2^{j-1}}{4^j\left(j!\right)^2}\sigma^{2j}\,d\sigma ds\\
		&=\frac{c_1}{3}r^2+\frac{c_2}{r}\int_0^r\sum_{j=1}^k\frac{c_1c_2^{j-1}}{4^j\left(j!\right)^2\left(2j+1\right)}s^{2j+2}\,ds\\
		&=\frac{c_1}{3}r^2+\sum_{j=1}^k\frac{c_1c_2^j}{\left(1-\frac{1}{4\left(j+1\right)^2}\right)4^{j+1}\left(\left(j+1\right)!\right)^2}r^{2j+2}=\sum_{j=1}^{k+1}\frac{c_1c_2^{j-1}}{\left(1-\frac{1}{4j^2}\right)4^j\left(j!\right)^2}r^{2j}.
		\end{align*}
	\end{proof}
\end{lma}
We can now prove the following result:
\begin{thm}
	Let \cref{cond:externalpotential,cond:ansatz} hold. Then, $\left(\left(\phi^k,A_\varphi^k,A_3^k\right)\right)_{k\in\N_0}$, where
	\begin{align*}
	\left(\phi^0,A_\varphi^0,A_3^0\right)=\left(0,0,0\right),\quad\left(\phi^{k+1},A_\varphi^{k+1},A_3^{k+1}\right)=\M\left(\phi^k,A_\varphi^k,A_3^k\right),\,k\in\N_0,
	\end{align*}
	is a Cauchy sequence in $C\left(\left[0,R_0\right];\R^3\right)$. The limit $\left(\phi,A_\varphi,A_3\right)$ is the fixed point of $\M$, whence $\left(\left(\f\right)_\alpha,\phi,A\right)$ is an axially symmetric steady state of the two and one-half dimensional relativistic Vlasov-Maxwell system on $\overline\Omega$ with external potential $A^\ext$, where the $\f$ are written in terms of $\phi$ and $A$, cf. \cref{eq:ansatzex}.
	\begin{proof}
		We abbreviate $P^k\coloneqq\left(\phi^k,A_\varphi^k,A_3^k\right)$ for $k\in\N_0$. By \cref{lma:aprioriiterates}, \cref{eq:gLipschitz}, and $0\leq\sigma\leq s\leq r$ we have
		\begin{align*}
		\left|\phi^{k+1}\left(r\right)-\phi^k\left(r\right)\right|,\left|A_\varphi^{k+1}\left(r\right)-A_\varphi^k\left(r\right)\right|,\left|A_3^{k+1}\left(r\right)-A_3^k\left(r\right)\right|\leq C\left(S\right)\int_0^r\int_0^s\left|P^k\left(\sigma\right)-P^{k-1}\left(\sigma\right)\right|\,d\sigma ds
		\end{align*}
		and thus
		\begin{align*}
		\left|P^{k+1}\left(r\right)-P^k\left(r\right)\right|\leq \sqrt{3}C\left(S\right)\int_0^r\int_0^s\left|P^k\left(\sigma\right)-P^{k-1}\left(\sigma\right)\right|\,d\sigma ds
		\end{align*}
		for $r\in\left[0,R_0\right]$, $k\in\N$. With $C\coloneqq\sqrt{3}C\left(S\right)$ this yields
		\begin{align*}
		\left|P^{k+1}\left(r\right)-P^k\left(r\right)\right|\leq\frac{SC^k}{\left(2k\right)!}r^{2k}
		\end{align*}
		for each $r\in\left[0,R_0\right]$, $k\in\N_0$ via induction: Indeed, this estimate obviously holds true for $k=0$, and moreover we have
		\begin{align*}
		\left|P^{k+1}\left(r\right)-P^k\left(r\right)\right|&\leq C\int_0^r\int_0^s\left|P^k\left(\sigma\right)-P^{k-1}\left(\sigma\right)\right|\,d\sigma ds\leq C\int_0^r\int_0^s\frac{SC^{k-1}}{\left(2k-2\right)!}\sigma^{2k-2}\,d\sigma ds\\
		&=\frac{SC^k}{\left(2k-1\right)!}\int_0^rs^{2k-1}\,ds=\frac{SC^k}{\left(2k\right)!}r^{2k}
		\end{align*}
		for $k\geq 1$. Therefore, for each $m\geq k$ and $r\in\left[0,R_0\right]$ there holds
		\begin{align*}
		\left|P^m\left(r\right)-P^k\left(r\right)\right|\leq\sum_{j=k}^{m-1}\left|P^{j+1}\left(r\right)-P^j\left(r\right)\right|\leq\sum_{j=k}^{m-1}\frac{SC^k}{\left(2k\right)!}r^{2k}\leq\sum_{j=k}^\infty\frac{SC^k}{\left(2k\right)!}R_0^{2k}.
		\end{align*}
		Since the series $\displaystyle\sum_{j=0}^\infty\frac{SC^k}{\left(2k\right)!}R_0^{2k}$ converges, it follows that $\left(P^k\right)$ is a Cauchy sequence in the space $C\left(\left[0,R_0\right];\R^3\right)$. Passing to the limit, we easily see that
		\begin{align*}
		\left(\phi,A_\varphi,A_3\right)=\lim_{k\to\infty}\left(\phi^{k+1},A_\varphi^{k+1},A_3^{k+1}\right)=\lim_{k\to\infty}\M\left(\phi^k,A_\varphi^k,A_3^k\right)=\M\left(\phi,A_\varphi,A_3\right),
		\end{align*}
		since $\M$ is continuous due to \cref{lma:Mcontcomp}. Hence, $\left(\phi,A_\varphi,A_3\right)$ is a (and by \cref{thm:MfixedpointSteadyState} the) fixed point of $\M$ and the corresponding tuple $\left(\left(\f\right)_\alpha,\phi,A\right)$ is a steady state.
	\end{proof}
\end{thm}

\subsection{Further properties}
A desirable property of a steady state is that it is compactly supported with respect to $v$. It is well known in similar settings that a necessary and sufficient condition for this is that there exists a cut-off energy. Indeed, the existence of such a cut-off energy guarantees this property also in our setting, as is shown below. Another obvious property which should hold is that the steady state is nontrivial -- for example, we have not excluded the pointless possibility $\etaa=0$ yet. We first state conditions under which a steady state indeed has these two properties and then prove the corresponding theorem.
\begin{cond}
	For each $\alpha=1,\dots,N$ there holds:
	\begin{condlist}\label{cond:compsuppnontrivial}
		\item\label{cond:compsupp} there exists $\E_0\in\R$ such that $\etaa\left(\mathcal E,\mathcal F,\mathcal G\right)=0$ if $\mathcal E\geq\E_0$;
		\item\label{cond:nontrivial} there exist $\E_u>\m$, $\G_l<0$, $\G_u>0$, and
		\begin{condoptionlist}
			\item\label{cond:nontrivialoption1} $\F_l<0$, $\F_u\geq 0$ or
			\item\label{cond:nontrivialoption2} $\F_l\leq 0$, $\F_u>0$
		\end{condoptionlist}
		such that
		\begin{align*}
		\forall\,\left(\mathcal E,\mathcal F,\mathcal G\right)\in\left]\m,\E_u\right[\times\left]\F_l,\F_u\right[\times\left]\G_l,\G_u\right[:\etaa\left(\mathcal E,\mathcal F,\mathcal G\right)>0.
		\end{align*}
	\end{condlist}
\end{cond} 
\begin{thm}\label{thm:compsuppnontrivial}
	Let \cref{cond:externalpotential,cond:ansatz} hold and let $\left(\left(\f\right)_\alpha,\phi,A\right)$ be a steady state, where $\left(\phi,A_\varphi,A_3\right)$ is the fixed point of $\M$ and the $\f$ are given by \cref{eq:ansatzex}. Then we have:
	\begin{thmlist}
		\item\label{thm:compsupp} If \cref{cond:compsupp} is satisfied, then the steady state is compactly supported with respect to $v$.
		\item\label{thm:nontrivial} If \cref{cond:nontrivial} is satisfied, then the steady state is nontrivial.
	\end{thmlist}
	\begin{proof}
		As for \ref{thm:compsupp}, we find that, if
		\begin{align*}
		\left|v\right|\geq\max_{\alpha=1,\dots,N}\left(\E_0+\left|\e\right|\xi\left(R_0\right)\right),
		\end{align*}
		then for each $\alpha=1,\dots,N$ and $x\in\overline\Omega$ there holds
		\begin{align*}
		\E\left(x,v\right)=\vo+\e\phi\left(r\right)\geq\left|v\right|-\left|\e\right|\xi\left(R_0\right)\geq\E_0
		\end{align*}
		due to \cref{lma:apriori} and hence $\f\left(x,v\right)=0$.
		
		As for \ref{thm:nontrivial}, we follow the idea of \cite{Kno19}. For fixed $\alpha\in\left\{1,\dots,N\right\}$ choose $0<r_\alpha\leq\frac{R_0}{2}$ small enough such that
		\begin{gather*}
		\sqrt{\m^2+2r_\alpha}-\left|\e\right|\xi\left(2r_\alpha\right)>\m,\quad\sqrt{\m^2+8r_\alpha}+\left|\e\right|\xi\left(2r_\alpha\right)<\E_u,\\
		\sqrt{r_\alpha}+\left|\e\right|\xi\left(2r_\alpha\right)+\left|\e\right|\sup_{0\leq r\leq 2r_\alpha}\left|A_3^\ext\left(r\right)\right|<\max\left\{-\G_l,\G_u\right\}
		\end{gather*}
		and
		\begin{align*}
		4r_\alpha^{\frac{3}{2}}+2\left|\e\right|r_\alpha\zeta\left(r_\alpha\right)+2\left|\e\right|r_\alpha\sup_{0\leq r\leq 2r_\alpha}\left|A_\varphi^\ext\left(r\right)\right|&<-\F_l,\\
		-\frac{1}{\sqrt{2}}r_\alpha^{\frac{3}{2}}+2\left|\e\right|r_\alpha\zeta\left(r_\alpha\right)+2\left|\e\right|r_\alpha\sup_{0\leq r\leq 2r_\alpha}\left|A_\varphi^\ext\left(r\right)\right|&<0
		\end{align*}
		in \cref{cond:nontrivialoption1} and
		\begin{align*}
		4r_\alpha^{\frac{3}{2}}+2\left|\e\right|r_\alpha\zeta\left(r_\alpha\right)+2\left|\e\right|r_\alpha\sup_{0\leq r\leq 2r_\alpha}\left|A_\varphi^\ext\left(r\right)\right|&<\F_u,\\
		\frac{1}{\sqrt{2}}r_\alpha^{\frac{3}{2}}-2\left|\e\right|r_\alpha\zeta\left(r_\alpha\right)-2\left|\e\right|r_\alpha\sup_{0\leq r\leq 2r_\alpha}\left|A_\varphi^\ext\left(r\right)\right|&>0
		\end{align*}
		in \cref{cond:nontrivialoption2}, respectively. Indeed, this choice of $r_\alpha$ is possible, since there holds $\xi\left(r\right),\zeta\left(r\right),rA_\varphi^\ext\left(r\right)=\mathcal O\left(r^2\right)$ for $r\to 0$, $A_3^\ext\left(0\right)=0$, and $\frac{1}{2},\frac{3}{2}\in\left]0,2\right[$. Next, let $\theta_\alpha\coloneqq\frac{3\pi}{2}$ in \cref{cond:nontrivialoption1} and $\theta_\alpha\coloneqq\frac{\pi}{2}$ in \cref{cond:nontrivialoption2}, respectively, and let
		\begin{align*}
		S_\alpha\coloneqq\left\{\vphantom{\frac{\pi}{4}}\left(r,u,\theta,v_3\right)\in\left[0,R_0\right]\times\left[0,\infty\right[\times\left[0,2\pi\right]\times\R\mid r_\alpha<r<2r_\alpha,\sqrt{r_\alpha}<u<2\sqrt{r_\alpha},\phantom{\big\}.}\right.\\
		\omit\hfill$\displaystyle\left.\theta_\alpha-\frac{\pi}{4}<\theta<\theta_\alpha+\frac{\pi}{4},-\sqrt{r_\alpha}<v_3<\sqrt{r_\alpha}\right\}.$
		\end{align*}
		In $\left(r,u,\theta,v_3\right)$-coordinates, where is the polar angle in the $\left(v_1,v_2\right)$-plane with basis $\left(e_r,e_\varphi\right)$, there holds
		\begin{align*}
		\E\left(r,u,\theta,v_3\right)&=\sqrt{\m^2+u^2+v_3^2}+\e\phi\left(r\right),\\
		\F\left(r,u,\theta,v_3\right)&=r\left(u\sin\theta+\e A_\varphi\left(r\right)+\e A_\varphi^\ext\left(r\right)\right),\\
		\G\left(r,u,\theta,v_3\right)&=v_3+\e A_3\left(r\right)+\e A_3^\ext\left(r\right).
		\end{align*}
		For each $\left(r,u,\theta,v_3\right)\in S_\alpha$, we have by \cref{lma:apriori}
		\begin{align*}
		\E\left(r,u,\theta,v_3\right)&\geq\sqrt{\m^2+2r_\alpha}-\left|\e\right|\xi\left(2r_\alpha\right)>\m,\\
		\E\left(r,u,\theta,v_3\right)&\leq\sqrt{\m^2+8r_\alpha}+\left|\e\right|\xi\left(2r_\alpha\right)<\E_u,\\
		\G\left(r,u,\theta,v_3\right)&\geq-\sqrt{r_\alpha}-\left|\e\right|\xi\left(2r_\alpha\right)-\left|\e\right|\sup_{0\leq r\leq 2r_\alpha}\left|A_3^\ext\left(r\right)\right|>\G_l,\\
		\G\left(r,u,\theta,v_3\right)&\leq\sqrt{r_\alpha}\left|\e\right|\xi\left(2r_\alpha\right)\left|\e\right|\sup_{0\leq r\leq 2r_\alpha}\left|A_3^\ext\left(r\right)\right|<\G_u
		\end{align*}
		and
		\begin{align*}
		\F\left(r,u,\theta,v_3\right)&\geq -4r_\alpha^{\frac{3}{2}}-2\left|\e\right|r_\alpha\zeta\left(r_\alpha\right)-2\left|\e\right|r_\alpha\sup_{0\leq r\leq 2r_\alpha}\left|A_\varphi^\ext\left(r\right)\right|>\F_l,\\
		\F\left(r,u,\theta,v_3\right)&\leq-\frac{1}{\sqrt{2}}r_\alpha^{\frac{3}{2}}+2\left|\e\right|r_\alpha\zeta\left(r_\alpha\right)+2\left|\e\right|r_\alpha\sup_{0\leq r\leq 2r_\alpha}\left|A_\varphi^\ext\left(r\right)\right|<0\leq\F_u
		\end{align*}
		in \cref{cond:nontrivialoption1} and
		\begin{align*}
		\F\left(r,u,\theta,v_3\right)&\leq 4r_\alpha^{\frac{3}{2}}+2\left|\e\right|r_\alpha\zeta\left(r_\alpha\right)+2\left|\e\right|r_\alpha\sup_{0\leq r\leq 2r_\alpha}\left|A_\varphi^\ext\left(r\right)\right|<\F_u,\\
		\F\left(r,u,\theta,v_3\right)&\geq\frac{1}{\sqrt{2}}r_\alpha^{\frac{3}{2}}-2\left|\e\right|r_\alpha\zeta\left(r_\alpha\right)-2\left|\e\right|r_\alpha\sup_{0\leq r\leq 2r_\alpha}\left|A_\varphi^\ext\left(r\right)\right|>0\geq\F_l
		\end{align*}
		in \cref{cond:nontrivialoption2}, respectively. Therefore,
		\begin{align*}
		ru\etaa\left(\E\left(r,u,\theta,v_3\right),\F\left(r,u,\theta,v_3\right),\G\left(r,u,\theta,v_3\right)\right)>0.
		\end{align*}
		Thus, we have (cf. proof of \cref{lma:reprdensities})
		\begin{align*}
		\int_{B_{R_0}}\int_{\R^3}\f\,dvd\left(x_1,x_2\right)&=2\pi\int_0^{R_0}r\int_{\R^3}\etaa\left(\E,\F,\G\right)\,dvdr\\
		&=2\pi\int_0^{R_0}\int_\R\int_0^\infty\int_0^{2\pi}ru\etaa\left(\E,\F,\G\right)\,d\theta dudv_3dr\\
		&\geq\int_{S_\alpha}ru\etaa\left(\E,\F,\G\right)\,d\left(r,u,\theta,v_3\right)>0,
		\end{align*}
		since $S_\alpha$ has positive Lebesgue measure. In particular, $\f\not\equiv 0$.
	\end{proof}
\end{thm}
\begin{rem}
	Vividly, the proof of \cref{thm:nontrivial} shows that, for each species, there are some particles near the symmetry axis with small momentum. Moreover, it was proved that in \cref{cond:nontrivialoption1} (or \ref{cond:nontrivialoption2}, respectively) there are some particles with negative (or positive, respectively) canonical angular momentum.
\end{rem}

\section{Confined steady states}\label{sec:confined}
There remains to find conditions on the external potential $A^\ext$ and the ansatz functions $\etaa$ under which a corresponding steady state is confined. We consider two possibilities:
\begin{itemize}
	\item A suitable $A_\varphi^\ext$ (corresponding to an external magnetic field in the $e_3$-direction) ensures confinement. This configuration is often called \enquote{$\theta$-pinch}.
	\item A suitable $A_3^\ext$ (corresponding to an external magnetic field in the $e_\varphi$-direction) ensures confinement. This configuration is often called \enquote{$z$-pinch}.
\end{itemize}
A combination of these two -- often called \enquote{screw-pinch} -- would of course also be possible, whence the following options are not exhaustive:
\begin{thm}
	Let \cref{cond:externalpotential,cond:ansatz,cond:compsuppnontrivial} hold and let $\left(\left(\f\right)_\alpha,\phi,A\right)$ be a steady state, where $\left(\phi,A_\varphi,A_3\right)$ is the fixed point of $\M$ and the $\f$ are given by \cref{eq:ansatzex}. We define
	\begin{align*}
	\mathcal N\coloneqq\left\{\alpha\in\left\{1,\dots,N\right\}\mid\e<0\right\},\quad\mathcal P\coloneqq\left\{\alpha\in\left\{1,\dots,N\right\}\mid\e>0\right\}.
	\end{align*}
	Furthermore, let $0<R<R_0$ and one of the following options hold:
	\begin{thmlist}
		\item ($\theta$-pinch)
		\begin{thmoptionlist}
			\item\label{thm:thetapinchoptiona} For each $\alpha\in\mathcal N$, \cref{cond:nontrivialoption1} is satisfied and we have $\etaa\left(\mathcal E,\mathcal F,\mathcal G\right)=0$ whenever $\mathcal F\geq 0$ (thus, necessarily $\F_u=0$). For each $\alpha\in\mathcal P$, \cref{cond:nontrivialoption2} is satisfied and we have $\etaa\left(\mathcal E,\mathcal F,\mathcal G\right)=0$ whenever $\mathcal F\leq 0$ (thus, necessarily $\F_l=0$). Moreover, assume
			\begin{align*}
			A_\varphi^\ext\left(r\right)\leq -a_\varphi\left(r\right),\quad R\leq r\leq R_0.
			\end{align*}
			\item\label{thm:thetapinchoptionb} For each $\alpha\in\mathcal N$, \cref{cond:nontrivialoption2} is satisfied and we have $\etaa\left(\mathcal E,\mathcal F,\mathcal G\right)=0$ whenever $\mathcal F\leq 0$ (thus, necessarily $\F_l=0$). For each $\alpha\in\mathcal P$, \cref{cond:nontrivialoption1} is satisfied and we have $\etaa\left(\mathcal E,\mathcal F,\mathcal G\right)=0$ whenever $\mathcal F\geq 0$ (thus, necessarily $\F_u=0$). Moreover, assume
			\begin{align*}
			A_\varphi^\ext\left(r\right)\geq a_\varphi\left(r\right),\quad R\leq r\leq R_0.
			\end{align*}
		\end{thmoptionlist}
		Here,
		\begin{align*}
		a_\varphi\left(r\right)\coloneqq\max_{\alpha=1,\dots,N}\frac{\sqrt{\left(\E_0+\left|\e\right|\xi\left(r\right)\right)^2-\m^2}}{\left|\e\right|}+\zeta\left(r\right).
		\end{align*}
		\item ($z$-pinch)
		\begin{thmoptionlist}
			\item\label{thm:zpinchoptiona} For each $\alpha\in\mathcal N$, there exists $\G_0<0$ such that $\etaa\left(\mathcal E,\mathcal F,\mathcal G\right)=0$ whenever $\mathcal G\leq\G_0$. For each $\alpha\in\mathcal P$, there exists $\G_0>0$ such that $\etaa\left(\mathcal E,\mathcal F,\mathcal G\right)=0$ whenever $\mathcal G\geq\G_0$. Moreover, assume
			\begin{align*}
			A_3^\ext\left(r\right)\geq a_3\left(r\right),\quad R\leq r\leq R_0.
			\end{align*}
			\item\label{thm:zpinchoptionb} For each $\alpha\in\mathcal N$, there exists $\G_0>0$ such that $\etaa\left(\mathcal E,\mathcal F,\mathcal G\right)=0$ whenever $\mathcal G\geq\G_0$. For each $\alpha\in\mathcal P$, there exists $\G_0<0$ such that $\etaa\left(\mathcal E,\mathcal F,\mathcal G\right)=0$ whenever $\mathcal G\leq\G_0$. Moreover, assume
			\begin{align*}
			A_3^\ext\left(r\right)\leq -a_3\left(r\right),\quad R\leq r\leq R_0.
			\end{align*}
		\end{thmoptionlist}
		Here,
		\begin{align*}
		a_3\left(r\right)\coloneqq\max_{\alpha=1,\dots,N}\frac{\left|\G_0\right|+\sqrt{\left(\E_0+\left|\e\right|\xi\left(r\right)\right)^2-\m^2}}{\left|\e\right|}+\xi\left(r\right).
		\end{align*}
	\end{thmlist}
	Then, the steady state is confined with radius at most $R$, compactly supported with respect to $v$, and nontrivial.
	\begin{proof}
		First note that for each $\left(x,v\right)\in\overline\Omega\times\R^3$ and $\alpha=1,\dots,N$ we have $\f\left(x,v\right)=0$ if
		\begin{align*}
		\left|v\right|\geq\sqrt{\left(\E_0+\left|\e\right|\xi\left(r\right)\right)^2-\m^2},
		\end{align*}
		since then
		\begin{align*}
		\E\left(x,v\right)\geq\sqrt{\m^2+\left|v\right|^2}-\left|\e\right|\xi\left(r\right)\geq\E_0
		\end{align*}
		by \cref{lma:apriori}. Thus, for each $\alpha=1,\dots,N$ it suffices to consider $v\in\R^3$ with
		\begin{align*}
		\left|v\right|<\sqrt{\left(\E_0+\left|\e\right|\xi\left(r\right)\right)^2-\m^2}.
		\end{align*}
		In the following, always let $r\in\left[R,R_0\right]$, $\alpha\in\mathcal N$, $\beta\in\mathcal P$, and $v$ as above.
		
		If \cref{thm:thetapinchoptiona} is satisfied, there holds
		\begin{align*}
		\F\left(x,v\right)&\geq r\left(-\left|v\right|+\e\zeta\left(r\right)+\e A_\varphi^\ext\left(r\right)\right)\geq r\left(-\left|v\right|+\e\zeta\left(r\right)-\e a_\varphi\left(r\right)\right)\\
		&\geq r\left(-\sqrt{\left(\E_0+\left|\e\right|\xi\left(r\right)\right)^2-\m^2}+\e\zeta\left(r\right)-\e\left(\frac{\sqrt{\left(\E_0+\left|\e\right|\xi\left(r\right)\right)^2-\m^2}}{-\e}+\zeta\left(r\right)\right)\right)=0,\\
		\mathcal F^\beta\left(x,v\right)&\leq r\left(\left|v\right|+q_\beta\zeta\left(r\right)+q_\beta A_\varphi^\ext\left(r\right)\right)\leq r\left(\left|v\right|+q_\beta\zeta\left(r\right)-q_\beta a_\varphi\left(r\right)\right)\\
		&\leq r\left(\sqrt{\left(\mathcal E_0^\beta+\left|q_\beta\right|\xi\left(r\right)\right)^2-m_\beta^2}+q_\beta\zeta\left(r\right)-q_\beta\left(\frac{\sqrt{\left(\mathcal E_0^\beta+\left|q_\beta\right|\xi\left(r\right)\right)^2-m_\beta^2}}{q_\beta}+\zeta\left(r\right)\right)\right)=0
		\end{align*}
		and thus $\f\left(x,v\right)=f^\beta\left(x,v\right)=0$.
		
		If \cref{thm:thetapinchoptionb} is satisfied, there holds
		\begin{align*}
		\F\left(x,v\right)&\leq r\left(\left|v\right|-\e\zeta\left(r\right)+\e A_\varphi^\ext\left(r\right)\right)\leq r\left(\left|v\right|-\e\zeta\left(r\right)+\e a_\varphi\left(r\right)\right)\\
		&\leq r\left(\sqrt{\left(\E_0+\left|\e\right|\xi\left(r\right)\right)^2-\m^2}-\e\zeta\left(r\right)+\e\left(\frac{\sqrt{\left(\E_0+\left|\e\right|\xi\left(r\right)\right)^2-\m^2}}{-\e}+\zeta\left(r\right)\right)\right)=0,\\
		\mathcal F^\beta\left(x,v\right)&\geq r\left(-\left|v\right|-q_\beta\zeta\left(r\right)+q_\beta A_\varphi^\ext\left(r\right)\right)\geq r\left(-\left|v\right|-q_\beta\zeta\left(r\right)+q_\beta a_\varphi\left(r\right)\right)\\
		&\geq r\left(-\sqrt{\left(\mathcal E_0^\beta+\left|q_\beta\right|\xi\left(r\right)\right)^2-m_\beta^2}-q_\beta\zeta\left(r\right)+q_\beta\left(\frac{\sqrt{\left(\mathcal E_0^\beta+\left|q_\beta\right|\xi\left(r\right)\right)^2-m_\beta^2}}{q_\beta}+\zeta\left(r\right)\right)\right)=0
		\end{align*}
		and thus $\f\left(x,v\right)=f^\beta\left(x,v\right)=0$.
		
		If \cref{thm:zpinchoptiona} is satisfied, there holds
		\begin{align*}
		\G&\left(x,v\right)\leq\left|v\right|-\e\xi\left(r\right)+\e A_3^\ext\left(r\right)\leq\left|v\right|-\e\xi\left(r\right)+\e a_3\left(r\right)\\
		&\leq\sqrt{\left(\E_0+\left|\e\right|\xi\left(r\right)\right)^2-\m^2}-\e\xi\left(r\right)+\e\left(\frac{-\G_0+\sqrt{\left(\E_0+\left|\e\right|\xi\left(r\right)\right)^2-\m^2}}{-\e}+\xi\left(r\right)\right)=\G_0,\\
		\mathcal G^\beta&\left(x,v\right)\geq-\left|v\right|-q_\beta\xi\left(r\right)+q_\beta A_3^\ext\left(r\right)\geq-\left|v\right|-q_\beta\xi\left(r\right)+q_\beta a_3\left(r\right)\\
		&\geq-\sqrt{\left(\mathcal E_0^\beta+\left|q_\beta\right|\xi\left(r\right)\right)^2-m_\beta^2}-q_\beta\xi\left(r\right)+q_\beta\left(\frac{\mathcal G_0^\beta+\sqrt{\left(\mathcal E_0^\beta+\left|q_\beta\right|\xi\left(r\right)\right)^2-m_\beta^2}}{q_\beta}+\xi\left(r\right)\right)=\mathcal G_0^\beta
		\end{align*}
		and thus $\f\left(x,v\right)=f^\beta\left(x,v\right)=0$.
		
		If \cref{thm:zpinchoptionb} is satisfied, there holds
		\begin{align*}
		\G&\left(x,v\right)\geq-\left|v\right|+\e\xi\left(r\right)+\e A_3^\ext\left(r\right)\geq-\left|v\right|+\e\xi\left(r\right)-\e a_3\left(r\right)\\
		&\geq-\sqrt{\left(\E_0+\left|\e\right|\xi\left(r\right)\right)^2-\m^2}+\e\xi\left(r\right)-\e\left(\frac{\G_0+\sqrt{\left(\E_0+\left|\e\right|\xi\left(r\right)\right)^2-\m^2}}{-\e}+\xi\left(r\right)\right)=\G_0,\\
		\mathcal G^\beta&\left(x,v\right)\leq\left|v\right|+q_\beta\xi\left(r\right)+q_\beta A_3^\ext\left(r\right)\leq\left|v\right|+q_\beta\xi\left(r\right)-q_\beta a_3\left(r\right)\\
		&\leq\sqrt{\left(\mathcal E_0^\beta+\left|q_\beta\right|\xi\left(r\right)\right)^2-m_\beta^2}+q_\beta\xi\left(r\right)-q_\beta\left(\frac{-\mathcal G_0^\beta+\sqrt{\left(\mathcal E_0^\beta+\left|q_\beta\right|\xi\left(r\right)\right)^2-m_\beta^2}}{q_\beta}+\xi\left(r\right)\right)=\mathcal G_0^\beta
		\end{align*}
		and thus $\f\left(x,v\right)=f^\beta\left(x,v\right)=0$.
		
		Hence, in all four cases the steady state is confined with radius at most $R$. That the steady state is compactly supported with respect to $v$ and nontrivial has already been proved in \cref{thm:compsuppnontrivial}.
	\end{proof}
\end{thm}
Vividly, for example \cref{thm:thetapinchoptiona} says that all negatively (positively) charged particles have negative (positive) canonical angular momentum thanks to the ansatz function and that, however, for $R\leq r\leq R_0$ a sufficiently small negative $A_\varphi^\ext$ would cause a positive (negative) canonical angular momentum of negatively (positively) charged particles possibly located there. Similarly, for example \cref{thm:zpinchoptiona} says that there cannot exist negatively (positively) charged particles with too small (large) third component of the canonical momentum thanks to the ansatz function and that, however, for $R\leq r\leq R_0$ a sufficiently large positive $A_3^\ext$ would cause a too small (large) third component of the canonical momentum of negatively (positively) charged particles possibly located there.

Since $A_\varphi^\ext\left(0\right)=A_3^\ext\left(0\right)=0$ due to \cref{cond:externalpotential} and $a_\varphi\left(0\right)\neq 0\neq a_3\left(0\right)$ due to \cref{cond:nontrivial}, $\left|A_\varphi^\ext\right|$ or $\left|A_3^\ext\right|$, respectively, has to increase sufficiently fast on $\left[0,R\right]$ to satisfy the respective condition on $\left[R,R_0\right]$. Moreover, $a_\varphi$ and $a_3$ increase when the ansatz functions $\etaa$ (and hence $\xi$, $\zeta$) increase. Thus, a larger external magnetic field is necessary to confine a larger amount of particles (as one would expect).

To obtain a specific example for an external magnetic field ensuring confinement, we consider a $\theta$-pinch configuration and a homogeneous external magnetic field parallel to the symmetry axis, i.e., $B^\ext=B_3^\ext e_3$ and $B_3^\ext\equiv b$ for some constant $b\in\R$. As $B_3^\ext\left(r\right)=\frac{1}{r}\left(rA_\varphi^\ext\left(r\right)\right)'$ and $A_\varphi^\ext\left(0\right)=0$, there has to hold $A_\varphi^\ext\left(r\right)=br$. Therefore, the steady state is confined for a sufficiently strong external magnetic field, that is to say if
\begin{align*}
\left|b\right|\geq\sup_{r\in\left[R,R_0\right]}\frac{a_\varphi\left(r\right)}{r}
\end{align*}
and $b<0$ (if \cref{thm:thetapinchoptiona} is satisfied) or $b>0$ (if \cref{thm:thetapinchoptionb} is satisfied), respectively. As opposed to this, no configuration can exist where the $\varphi$-component of the external magnetic field is constant (and nontrivial), since in this case $A_3^\ext$ would have to be a linear function of $r$ because of $B_\varphi^\ext=-\left(A_3^\ext\right)'$ and $A_3^\ext\left(0\right)=0$, which contradicts the necessary condition $\left(A_3^\ext\right)'\left(0\right)=0$.

We finish with an important remark:
\begin{rem}
	Another interesting setting is that there is no confinement device and thus no boundary at $r=R_0$ in the first place. In this case, $\Omega=\R^3$ and no boundary conditions at $r=R_0$ have to be imposed. Moreover, \cref{def:steadystate} can be suitably adapted to this new setting by abolishing \cref{eq:BCstat} and setting $R_0=\infty$. However, if we seek a steady state of this new setting that is confined with radius at most $R>0$, we firstly choose a (slightly) larger $R_0>R$, secondly consider the confinement problem as before with boundary at $r=R_0$ and choose $A_\varphi^\ext$ or $A_3^\ext$ suitably to ensure confinement of the obtained steady state with radius at most $R$, and thirdly \enquote{glue} this steady state defined on $\left[0,R_0\right]$ and the vacuum solution on $\left[R_0,\infty\right[$ together, i.e., extend each $\f$ by zero and the potentials by their respective integral formula, that is,
	\begin{align*}
	\phi\left(r\right)&=-4\pi\int_0^r\frac{1}{s}\int_0^s\sigma\rho\left(\sigma\right)\,d\sigma ds=-4\pi\int_0^R\frac{1}{s}\int_0^s\sigma\rho\left(\sigma\right)\,d\sigma ds-4\pi\int_R^r\frac{1}{s}\int_0^R\sigma\rho\left(\sigma\right)\,d\sigma ds\\
	&=-4\pi\int_0^R\frac{1}{s}\int_0^s\sigma\rho\left(\sigma\right)\,d\sigma ds-4\pi\int_0^Rs\rho\left(s\right)\,ds\cdot\left(\ln r-\ln R\right),\\
	A_\varphi\left(r\right)&=-\frac{4\pi}{r}\int_0^rs\int_0^sj_\varphi\left(\sigma\right)\,d\sigma ds=-\frac{4\pi}{r}\int_0^Rs\int_0^sj_\varphi\left(\sigma\right)\,d\sigma ds-\frac{4\pi}{r}\int_R^rs\int_0^Rj_\varphi\left(\sigma\right)\,d\sigma ds\\
	&=-\frac{4\pi}{r}\int_0^Rs\int_0^sj_\varphi\left(\sigma\right)\,d\sigma ds-2\pi\int_0^R j_\varphi\left(s\right)\,ds\cdot\left(r-\frac{R^2}{r}\right),\\
	A_3\left(r\right)&=-4\pi\int_0^r\frac{1}{s}\int_0^s\sigma j_3\left(\sigma\right)\,d\sigma ds=-4\pi\int_0^R\frac{1}{s}\int_0^s\sigma j_3\left(\sigma\right)\,d\sigma ds-4\pi\int_R^r\frac{1}{s}\int_0^R\sigma j_3\left(\sigma\right)\,d\sigma ds\\
	&=-4\pi\int_0^R\frac{1}{s}\int_0^s\sigma j_3\left(\sigma\right)\,d\sigma ds-4\pi\int_0^Rs j_3\left(s\right)\,ds\cdot\left(\ln r-\ln R\right)
	\end{align*}
	for $r\geq R$. Note that for this procedure it is important that the $\f$ already vanish on $\left[R,R_0\right]$ so that the composite $\f$ have no jumps at $r=R_0$. With the identities above we can furthermore determine the asymptotics of the potentials for $r\to\infty$. In particular,
	\begin{gather*}
	\phi\left(r\right)=-4\pi a\ln r+\text{const.},\quad A_3\left(r\right)=-4\pi b\ln r+\text{const.},\quad r\geq R,\\
	\lim_{r\to\infty}\left(A_\varphi\left(r\right)+2\pi cr\right)=0,
	\end{gather*}
	where
	\begin{align*}
	a=\int_0^Rs\rho\left(s\right)\,ds,\quad b=\int_0^Rs j_3\left(s\right)\,ds,\quad c=\int_0^R j_\varphi\left(s\right)\,ds.
	\end{align*}
	Here, $a$ and $b$ can be interpreted as the total charge and the third component of the total current on each slice perpendicular to the symmetry axis.
\end{rem}

\nocite{*}
\bibliography{confststa}
\bibliographystyle{plain}
\end{document}